\documentclass[10pt,journal]{IEEEtran}

\usepackage{cite}
\usepackage[pdftex]{graphicx}
\usepackage[cmex10]{amsmath}
\usepackage{amssymb}
\usepackage{amsfonts}

\usepackage{algorithm}
\usepackage{algorithmic}
\usepackage[caption=false,font=footnotesize]{subfig}

\usepackage{color}
\usepackage{array}
\usepackage{mdwmath}
\usepackage{mdwtab}
\usepackage{eqparbox}

\begin{document}

\title{Cellular-Base-Station Assisted Device-to-Device Communications in TV White Space}

\author{Guoru~Ding,~\IEEEmembership{Member,~IEEE,} Jinlong~Wang,~\IEEEmembership{Senior Member,~IEEE}, Qihui~Wu,~\IEEEmembership{Senior Member,~IEEE,} Yu-Dong~Yao,~\IEEEmembership{Fellow,~IEEE,}~Fei Song, and Theodoros A. Tsiftsis,~\IEEEmembership{Senior Member,~IEEE}
\thanks{Manuscript received May 20, 2013, revised Oct. 14, 2013, accepted May 17, 2015.}
\thanks{G. Ding, J. Wang, Q. Wu, and F. Song are with the College of Communications Engineering, PLA University of Science and Technology, Nanjing, Jiangsu 210007, China. E-mail: dingguoru@gmail.com, wjl543@sina.com, wuqihui2014@sina.com, aroucian@163.com.}
\thanks{Y. Yao is with the Electrical and Computer Engineering Department, Stevens Institute of Technology, Hoboken, NJ 07030, USA. E-mail:yyao@stevens.edu.}
\thanks{T. A. Tsiftsis is with the Department of Electrical Engineering, Technological Educational Institute of Central Greek, 35100, Lamia, Greece. E-mail: tsiftsis@teiste.gr.}}

\maketitle

\begin{abstract}
This paper presents a systematic approach to exploit TV white space (TVWS) for device-to-device (D2D) communications with the aid of the existing cellular infrastructure. The goal is to build a \emph{location-specific} TVWS database, which provides a look-up table \emph{service} for any D2D link to determine its maximum permitted emission power (MPEP) in an unlicensed digital TV (DTV) band. To achieve this goal, the idea of mobile crowd sensing is firstly introduced to collect active spectrum measurements from massive personal mobile devices. Considering the incompleteness of crowd measurements, we formulate the problem of unknown measurements recovery as a matrix completion problem and apply a powerful fixed point continuation algorithm to reconstruct the unknown elements from the known elements. By joint exploitation of the big spectrum data in its vicinity, each cellular base station further implements a nonlinear support vector machine algorithm to perform \emph{irregular} coverage boundary detection of a licensed DTV transmitter. With the knowledge of the detected coverage boundary, an \emph{opportunistic} spatial reuse algorithm is developed for each D2D link to determine its MPEP. Simulation results show that the proposed approach can successfully enable D2D communications in TVWS while satisfying the interference constraint from the licensed DTV services. In addition, to our best knowledge, this is the first try to explore and exploit TVWS inside the DTV protection region resulted from the shadowing effect. Potential application scenarios include communications between internet of vehicles in the underground parking, D2D communications in hotspots such as subway, game stadiums, and airports, etc.
\end{abstract}


\IEEEpeerreviewmaketitle

\section{Introduction}
\IEEEPARstart{R}{ecently}, we have witnessed a dramatic growth in wireless data traffic. To provide higher capacity, researchers have been exploring the next generation wireless communication systems~\cite{JGA-5G,Niu,Mitola-5G}. However, one crucial impediment is the shortage of radio spectrum~\cite{Feifei,Yulong}. To develop efficient solutions, several approaches have been suggested, e.g., spectrum extension, network densification, and spectrum efficiency improvement, etc~\cite{whitepaper_5G}.

The integration of device-to-device (D2D) communications into cellular networks is a promising paradigm to improve spectrum efficiency. D2D communication enables mobile devices in proximity to establish a direct link and to reuse the cellular spectrum. The advantages are manifold: offloading cellular traffic, eliminating coverage holes, improving spatial reuse, reducing battery consumption, and thereby enabling new services~\cite{D2D_2008,D2D_2009,D2D_Mag,D2D_WMag}. Provided that D2D communications share the same \emph{licensed} spectrum resource with the co-located cellular users, majority of previous studies have focused on addressing the mutual interference between them~\cite{D2D_Mutual_Inter2,D2D_Mutual_Inter3,D2D_SLY,D2D_App}.

Differently, in this paper we focus on exploring \emph{unlicensed} TV white space (TVWS) for D2D communications in cellular networks. TVWS refers to TV bands that are not used by any DTV services at a particular time in a particular geographic area~\cite{What_2009}. TVWS is also known as \emph{digital dividend}, which emerges from the transition from analog to digital TV transmission\cite{Digital_Dividend}. With the transition, large portions of the UHF spectrum (e.g., 512-608 MHz and 614-698 MHz in USA, 470-550 MHz and 614-782 MHz in UK) have been released for unlicensed devices as long as they do not create harmful interference to any licensed services. The TVWS in the released UHF spectrum is quite attractive since it has favorable propagation and building penetration characteristics.

However, the development of D2D communications in TVWS poses critical challenges. First of all, strict interference constraint should be met to protect the normal operation of licensed DTV services~\cite{Inter_Ana2009,20dB_2006}. To develop efficient solutions, two main approaches have been suggested: i) spectrum sensing and ii) geolocation database. In the spectrum sensing approach, the availability of TVWS at a given location is modeled as a threshold-based hypothesis test~\cite{TWC_mine}. This approach, however, suffers from the hidden node problem because of shadowing~\cite{Hode_Node_Problem}. On the contrary, according to~\cite{FCC-Rule,FCC-TVWS,MicrosoftWhiteSpace,GoogleWhiteSpace,SpectrumBridgeWhiteSpace,Telcordia}, the geolocation database approach seems to provide a technically feasible and commercially viable solution in the near future. This approach provides a service that an unlicensed device can inquire the TVWS availability from a geolocation database, which predicts the availability of TVWS at any location using propagation modeling with high-resolution terrain data~\cite{Senseless_2012,Geo-Ding}. One key limitation of this approach is that the accuracy of the TVWS availability provided by the database depends highly on the quality of the propagation modeling and the granularity of the terrain data~\cite{Bounding,survey_PathLoss}.

Motivated by the opportunities and challenges described above, in this paper, we propose a novel mobile crowd sensing-driven geolocation database approach to explore TVWS for D2D communications. The goal is to build a \emph{location-specific} TVWS database, which provides a look-up table \emph{service} for any D2D link to determine its maximum permitted emission power (MPEP) in an unlicensed digital TV (DTV) band. Before presenting details of the proposed approach, some key design rationales are stated as follows:

(i) \emph{Using TVWS for D2D communications in cellular networks is promising in terms of both technical and commercial viability.} Just as a container filled with rocks still has room for sand, the geographical area not covered by large-scale DTV systems could be reused by small-scale D2D communications\footnote{The transmission range of a DTV transmitter can be up to hundreds of kilometers, while the transmission range of a D2D communication link in cellular networks can be several meters to hundreds of meters.} to improve the overall spatial spectrum utilization. Moreover, in comparison to other license-exempt spectral bands such as 2.4 GHz and 5 GHz, the superior propagation properties of TV spectrum allows a higher transmission range at much lower energy requirements. Furthermore, instead of introducing a new operator to invest in a new infrastructure, it can be a promising business model for the cellular service providers to exploit TVWS, since they can offer more bandwidth to mobile users/devices by accessing extra spectrum with an already-deployed cellular infrastructure.

(ii) \emph{There are much more spatial reuse opportunities for unlicensed devices in TV spectrum than we have already recognized.} After carefully studying the recent released DTV coverage maps by FCC~\cite{FCC-MAP} and Ofcom~\cite{Ofcom-MAP}, we observe that for a given DTV transmitter, there are plenty of spatial coverage holes resulted from the signal attenuation due to either distance-related path loss or shadowing effect of irregular terrain or buildings. Traditionally, shadowing has been considered as a physical barrier to reliable spectrum sensing because of the hidden node problem~\cite{Hode_Node_Problem}. However, from a perspective of spatial reuse, we see that shadowing can greatly improve the isolation between the small-scale D2D communications and large-scale DTV services, which thus increases the flexibility in local spectrum usage and benefits the D2D communications by providing more white spaces.

(iii) \emph{Parallel with the existing model-based approaches, data-driven approaches promise a new paradigm to explore TVWS availability.} The popularity of various mobile wireless devices (e.g., smart phones, tablets, and in-vehicle sensors) equipped with programmable and powerful sensors makes it a good potential to learn the TVWS availability from massive active spectrum measurements or big spectrum data, which can serve as an alternative of the current sophisticated propagation models in the geolocation database approach (see, e.g.,~\cite{Senseless_2012}).

Based on the design rationales above, the main contributions of this paper are summarized as follows:
\begin{itemize}
  \item Formulate the spatial reuse of a TV channel between licensed DTV services and unlicensed D2D communications as an optimization problem, where the permitted transmit power for an unlicensed mobile device is maximized, subject to i) a peak transmit power budget constrained by the hardware of that device, and ii) an interference probability threshold for the protection of licensed DTV receptions.
  \item Introduce the idea of mobile crowd sensing to collect massive spectrum measurements from personal mobile devices, with the aid of the existing cellular base stations (BS). Each cellular BS is in charge of collecting, pre-processing, and mining the measurement data of its cell and the neighboring cells to form a localized TVWS database service for D2D communications in its vicinity.
  \item Develop big spectrum data mining algorithms for each cellular BS to implement the task of TVWS geolocation database building: i) A fast matrix completion algorithm for obtaining complete spectrum status from a few known measurement samples, ii) a nonlinear support vector machine algorithm for performing irregular coverage boundary detection of a licensed DTV transmitter, and iii) an opportunistic spatial reuse algorithm for each D2D communication link to determine its MPEP.
  \item Provide in-depth simulations under two critical and representative scenarios, which show that compared with the state-of-the-art approach, the proposed approach has two main advantages: i) improved spatial reuse between large-scale DTV services and small-scale D2D communications can be obtained; ii) reduced interference to the potential DTV receptions could be achieved.
\end{itemize}

The reminder of this paper is as follows. Section II presents the system model. Section III presents the problem formulation and analysis. Section IV provides an overview of the proposed approach. Section V details the algorithm designs involved in the proposed approach. Section VI performs the performance evaluation and Section VII concludes the paper.

\begin{figure}[!t]
\centering
\includegraphics[width=\linewidth]{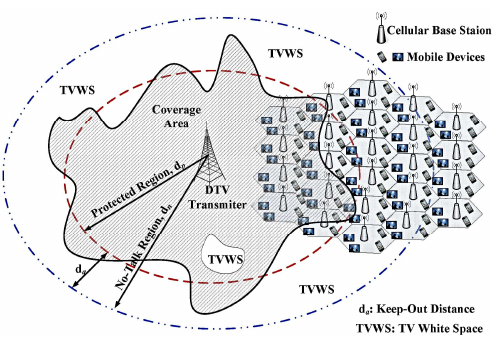}
\caption{An illustrative example of the network model used in this paper.}
\label{fig_Visio-DTV-CBS-D2D}
\end{figure}

\section{System Model}

\subsection{Network Model}
As shown in Fig. \ref{fig_Visio-DTV-CBS-D2D}, we consider a scenario that a DTV broadcasting network shares a TV channel with a cellular network. The high-power DTV transmitter has a spatial footprint up to hundreds of kilometers and its \emph{passive} DTV receivers can be located anywhere inside its coverage area. In Fig. \ref{fig_Visio-DTV-CBS-D2D}, the covered area of the DTV transmitter is shown in shade and the uncovered area or TVWS is shown in white. The irregular curve around the DTV transmitter represents the ground truth coverage boundary. The dashed circular curve denotes a commonly-used simplified coverage boundary with a protection region $d_p$. The dash-dotted circular curve represents the boundary of a no-talk-region~\cite{TVWS_Europe}, which is specified as the sum of the DTV protection region and an additional separation named keep-out distance that depends on the peak transmit power of unlicensed devices. The co-located cellular network consists of dozens of inter-connected cells. Each cell has a cellular BS located at the center and massive mobile devices (e.g., smart phones, tablets, and in-vehicle sensors) randomly distributed around it. The mobile devices can operate in two communication modes: i) communicate with or through the associated cellular BS, and ii) directly communicate to each other and bypass the cellular BS. The latter mode, also known as \emph{D2D communication}, is promising when devices are in close proximity to each other to provide higher throughput and enhance user experiences.

\subsection{Signal Propagation Model}
At a high level, a radio communication channel can be modeled as a multi-scale system with three major dynamics~\cite{survey_PathLoss}: large-scale distance-dependent path loss $L$, medium-scale shadowing $S$ due to fixed obstacles, and small-scale fading $F$ due to destructive interference from multipath effects and small scatterers. Denote $P_{{\bf{z}}}$ in dB as the transmit power of a transmitter located at position $\bf{z}$ and the received power at position $\bf{x}$ can be written as
\begin{align}
\label{eq:Propagation-model}
P_{{\bf{z}}\to{\bf{x}}}\texttt{[dB]}=P_{{\bf{z}}}-(L_{{\bf{z}}\to{\bf{x}}}+S_{{\bf{z}}\to{\bf{x}}}+F_{{\bf{z}}\to{\bf{x}}}).
\end{align}

For the propagation modeling of a wideband TV channel (e.g., 6 MHz in US and 8 MHz in UK), the small-scale fading $F$ are generally averaged out from multiple measurements as suggested in~\cite{Secure_2011}. Consequently, in the following, the main task of a signal propagation model is to predict the value of $L+S$ in Eq. (\ref{eq:Propagation-model}). Dozens of models have been proposed during the last 60 years~\cite{survey_PathLoss}, which can be generally grouped into two classes: i) deterministic propagation models\footnote{The Longley-Rice irregular terrain model (ITM) is one popular deterministic model~\cite{L-R}, which takes into account a wide variety of factors such as terrain shapes, climactic effects, soil conductivity, earth's curvature, etc~\cite{Senseless_2012}. One key issue of this model is the high computational complexity.}, and ii) statistical propagation models\footnote{The ITU-R model is a commonly-used statistical propagation model~\cite{TVWS_Europe}, in which all transmissions are assumed to be omnidirectional and the critical parameter is the distance to a DTV transmitter. The main drawback of this model is its inability to characterize the local variability in the received signal strength caused by irregular terrain effects.}.

In the following, we adopt a hybridized model, where
\begin{itemize}
  \item the\emph{ deterministic path loss component} is given as $L_{{\bf{z}}\to{\bf{x}}}=10\alpha \log_{10}(d_{{\bf{z}}\to{\bf{x}}})+20\log_{10}(f)+32.45$ as suggested in~\cite{Bounding}, where $\alpha$ is the path loss exponent depending on the specific propagation environment~\cite{TSP-WC-Book}, $d_{{\bf{z}}\to{\bf{x}}}$ in km is the distance between the transmitter and the receiver, and $f$ in MHz is the center frequency,
  \item and the \emph{random shadowing component} is modeled as a Gaussian variable $S_{{\bf{z}}  \to {\bf{x}}} \sim \mathcal{N}(\bar S_{{\bf{z}} \to {\bf{x}}}, \sigma _{S_{{\bf{z}}  \to {\bf{x}}} }^2)$. Notably, in this paper, we consider the case that the average shadowing loss $\bar S_{{\bf{z}}  \to {\bf{x}}}$ is \emph{environment (or location) dependent}, which can varies from 0 dB to tens of dBs~\cite{Hode_Node_Problem}. $\sigma _{S_{{\bf{z}}  \to {\bf{x}}} }$ is the shadow spread in dB.
\end{itemize}

\subsection{DTV Coverage Model}
Let $P_{\min}$ denote the minimum required power level that a DTV receiver can reliably decode the desired signal from its DTV transmitter. The \emph{location coverage probability} of a DTV receiver (located at $\bf{x}$) with respect to its DTV transmitter (located at $\bf{x_0}$) is defined as
\begin{align}
{\Pr}^{{\mathop{\rm cov}} } _{{\bf{x}}_0  \to {\bf{x}}}  &\buildrel \Delta \over= \Pr \{ P_{{\bf{x}}_0  \to {\bf{x}}}  \ge P_{\min } \} {\rm{= }}Q{\rm{(}}\frac{{P_{\min }  - \bar P_{{\bf{x}}_0  \to {\bf{x}}} }}{{\sigma _{S_{{\bf{x}}_0  \to {\bf{x}}} } }}{\rm{)}},
\end{align}
where $Q( \cdot )$ is the standard Gaussian tail function and $\bar P_{{\bf{x}}_0  \to {\bf{x}}}  =  P_{{\bf{x}_0} } - L_{{\bf{x}_0}\to{\bf{x}}}- \bar{S}_{{\bf{x}_0}\to {\bf{x}}}$ is the average received power at location $\bf{x}$.

\textbf{Definition 1}: For an interested area $\mathcal{A}$, denote $\nu _{{\mathop{\rm cov}}}$ as a location coverage threshold, the \emph{coverage area} of a DTV transmitter located at $\bf{x_0}$ can be defined as the set of all coverage locations, i.e.,
\begin{align}
\mathcal{A}_{\bf{x}_0}^{{\mathop{\rm cov}} } {\rm{ = \{  }}{\bf{x}}{\rm{|}}{\Pr}^{{\mathop{\rm cov}} } _{{\bf{x}}_0  \to {\bf{x}}} \ge \nu _{{\mathop{\rm cov}} } ,\forall {\bf{x}} \in \mathcal{A}\},
\end{align}
or equivalently,
\begin{align}
\mathcal{A}_{\bf{x}_0}^{{\mathop{\rm cov}} } {\rm{ = \{  }}{\bf{x}}{\rm{|}}\bar P_{{\bf{x}}_0  \to {\bf{x}}} \ge \bar P_{\min } ,\forall {\bf{x}} \in \mathcal{A}\},
\end{align}
where $\bar P_{\min } = P_{\min }  - \sigma _{s_{{\bf{x}}_0  \to {\bf{x}}} } Q^{ - 1} (\nu _{{\mathop{\rm cov}} } )$.

On the contrary, the area uncovered by the DTV transmitter located at $\bf{x_0}$, i.e., TV white space in $\mathcal{A}$, can be given as
\begin{align}
\mathcal{A}_{\bf{x}_0}^{{\mathop{\rm TVWS}} } {\rm{ = \{  }}{\bf{x}}{\rm{|}}\bar P_{{\bf{x}}_0  \to {\bf{x}}} < \bar P_{\min } ,\forall {\bf{x}} \in \mathcal{A}\}.
\end{align}

\emph{Remark 1:} Traditionally, to facilitate analysis, the DTV coverage area is usually simplified as a disc area with the DTV transmitter at the center. For instance, in~\cite{MV_2009}, the effects of shadowing component $S_{{\bf{x}}_0 \to {\bf{x}}}, \forall {\bf{x}} \in \mathcal{A}$, is neglected and only distance-related path loss is considered. In~\cite{BLMark_2009}, the shadowing component is modeled as a Gaussian variable while the mean $\bar S_{{\bf{x}}_0  \to {\bf{x}}}, \forall {\bf{x}} \in \mathcal{A}$ is assumed to be zero everywhere. However, in practice, the ground-truth coverage boundary of a DTV transmitter should be generally in irregular shape, resulted from the location-dependent shadowing variations in natural terrain morphology or man-made buildings. These shadowing variations can result in opportunities for TVWS exploitation that cannot be overlooked for the coexistence between small-scale D2D communications and large-scale DTV services.

\emph{Remark 2:} Compared with the existing disc models, the advantages of integrating location-dependent shadowing into the DTV coverage modeling are at least two folds: i) Improved spatial reuse can be obtained. Due to the effects of location-dependent shadowing, as shown in Fig. \ref{fig_Visio-DTV-CBS-D2D}, there can be some areas near the edge of or even inside the traditional disc coverage/protected region that are not covered by the DTV transmission. In those shadowing areas no DTV reception could work properly and thus spatial spectrum holes or TVWSs for unlicensed devices can be exploited. ii) Reduced interference to the DTV receptions near the edge of the ground-truth coverage area could be achieved. As illustrated in Fig. \ref{fig_Visio-DTV-CBS-D2D}, there are some areas covered by the DTV transmission while excluded by the traditional disc protected region, where the D2D communications will cause harmful interference to the potential DTV receptions.

\subsection{Unlicensed Device Interference Model}

Let $I_{\max}$ represent interference tolerance threshold of any DTV reception inside the DTV coverage area. The \emph{location interference probability} of an unlicensed device located at ${\bf{x}}_i$ with respect to a DTV receiver at ${\bf{x}} \in \mathcal{A}_{{\bf{x}}_0}^{{\mathop{\rm cov}}}$ is defined as
\begin{align}
{\Pr} _{{\bf{x}}_i  \to {\bf{x}}}^{{\mathop{\rm int}} }  &\buildrel \Delta \over = \Pr \{ P_{{\bf{x}}_i  \to {\bf{x}}}  \ge I_{\max } \} {\rm{ = }}Q{\rm{(}}\frac{{I_{\max }-\bar P_{{\bf{x}}_i  \to {\bf{x}}} }}{{\sigma _{S_{{\bf{x}}_i  \to {\bf{x}}} } }}{\rm{)}},
\end{align}
where $\bar P_{{\bf{x}}_i  \to {\bf{x}}}  = P_{{\bf{x}}_i} - L_{{\bf{x}_i}\to{\bf{x}}}- \bar{S}_{{\bf{x}}_i\to {\bf{x}}}$ is the average interference power perceived by the DTV receiver.

\textbf{Definition 2}: For an interested area $\mathcal{A}$, denote $\nu _{{\mathop{\rm int}}}$ as a location interference threshold and the \emph{interference area} of an unlicensed device located at ${\bf{x}}_i$ can be defined as the set of all interference locations, i.e.,
\begin{align}
\mathcal{A}^{{\mathop{\rm int}}}_{{\bf{x}}_i }  = \{ {\bf{x}}{\rm{|}} {\Pr} _{{\bf{x}}_i  \to {\bf{x}}}^{{\mathop{\rm int}} }  \ge \nu _{{\mathop{\rm int}} } ,\forall {\bf{x}} \in \mathcal{A}_{\bf{x}_0}^{{\mathop{\rm cov}}}\},
\end{align}
or equivalently,
\begin{align}
\mathcal{A}^{{\mathop{\rm int}}}_{{\bf{x}}_i } {\rm{ = \{  }}{\bf{x}}{\rm{|}}P_{{\bf{x}}_i }  \ge I_{\max ,{\bf{x}}_i  \to {\bf{x}}} ,\forall {\bf{x}} \in \mathcal{A}_{\bf{x}_0}^{{\mathop{\rm cov}} }\},
\end{align}
where $I_{\max ,{\bf{x}}_i  \to {\bf{x}}} \!=\! I_{\max } \! - \!\sigma _{S_{{\bf{x}}_i  \to {\bf{x}}} } Q^{ - 1} (\nu _{{\mathop{\rm int}} } )\!+ \!L_{{\bf{x}}_i\to{\bf{x}}}\!+\! \bar{S}_{{\bf{x}}_i  \to {\bf{x}}}$.

\emph{Remark 3:} Since the locations of DTV receivers are generally unknown \emph{a priori} to unlicensed devices, a conservative assumption should be adopted~\cite{Inter_Ana2009} that active DTV receptions can be located anywhere inside the DTV coverage area. Moreover, as shown in Fig.~\ref{fig_Visio-DTV-CBS-D2D}, in previous studies (see, e.g.,~\cite{TVWS_Europe}), for any unlicensed device, a \emph{no-talk-region} $d_n$ is specified as the sum of the DTV protection region $d_p$ and an additional keep-out distance $d_a$, which is the interference range that depends on the peak transmit power of that device. It is noted that the concept of no-talk-region simplifies the system design at the expense of significant TVWS loss, especially when efficient power control techniques can be used by unlicensed devices~\cite{TWC_mine}.

\section{Problem Formulation and Analysis}
Given the DTV coverage model and the unlicensed device interference model established above, the problem of interest in this paper is to determine whether a TVWS exists for a given D2D communication link in an interested area $\mathcal{A}$, and, if exists, its quantification. Technically, the problem can be expressed as: for a given unlicensed device located at position ${\bf{x}}_i  \in \mathcal{A}$, the objective is to maximize its transmit power $P_{{\bf{x}}_i }^*$, subject to i) the peak transmit power $P_{\texttt{peak}}$ constrained by its hardware and ii) the interference probability threshold $\nu _{{\mathop{\rm int}} }$ for protecting the licensed DTV receptions, i.e.,
\begin{align}
\textbf{\texttt{OP1}}:\quad P_{{\bf{x}}_i }^*  = \max {\rm{ }}P_{{\bf{x}}_i }
\end{align}
subject to
\begin{align}
\quad \quad \quad \quad \quad \quad \quad \quad &P_{{\bf{x}}_i }  \le {\rm{ }}P_{\texttt{peak} }, \\
\quad \quad \quad \quad \quad \quad \quad \quad \quad &{\Pr} _{{\bf{x}}_i  \to {\bf{x}}}^{{\mathop{\rm int}} }  \le \nu _{{\mathop{\rm int}} } ,\forall {\bf{x}} \in \mathcal{A}_{\bf{x}_0}^{{\mathop{\rm cov}} }.
\end{align}

Based on \textbf{Definition 2}, we can combine the two constraints in $\textbf{\texttt{OP1}}$ as follows
\begin{align}
\label{eq:Combination}
\quad \quad \quad \quad P_{{\bf{x}}_i }  \le \min \{ {\rm{ }}P_{\texttt{peak} } ,I_{\max ,{\bf{x}}_i  \to {\bf{x}}}, \forall {\bf{x}} \in \mathcal{A}_{\bf{x}_0}^{{\mathop{\rm cov}} }\}.
\end{align}

\begin{figure}[!t]
\centering
\includegraphics[width=\linewidth]{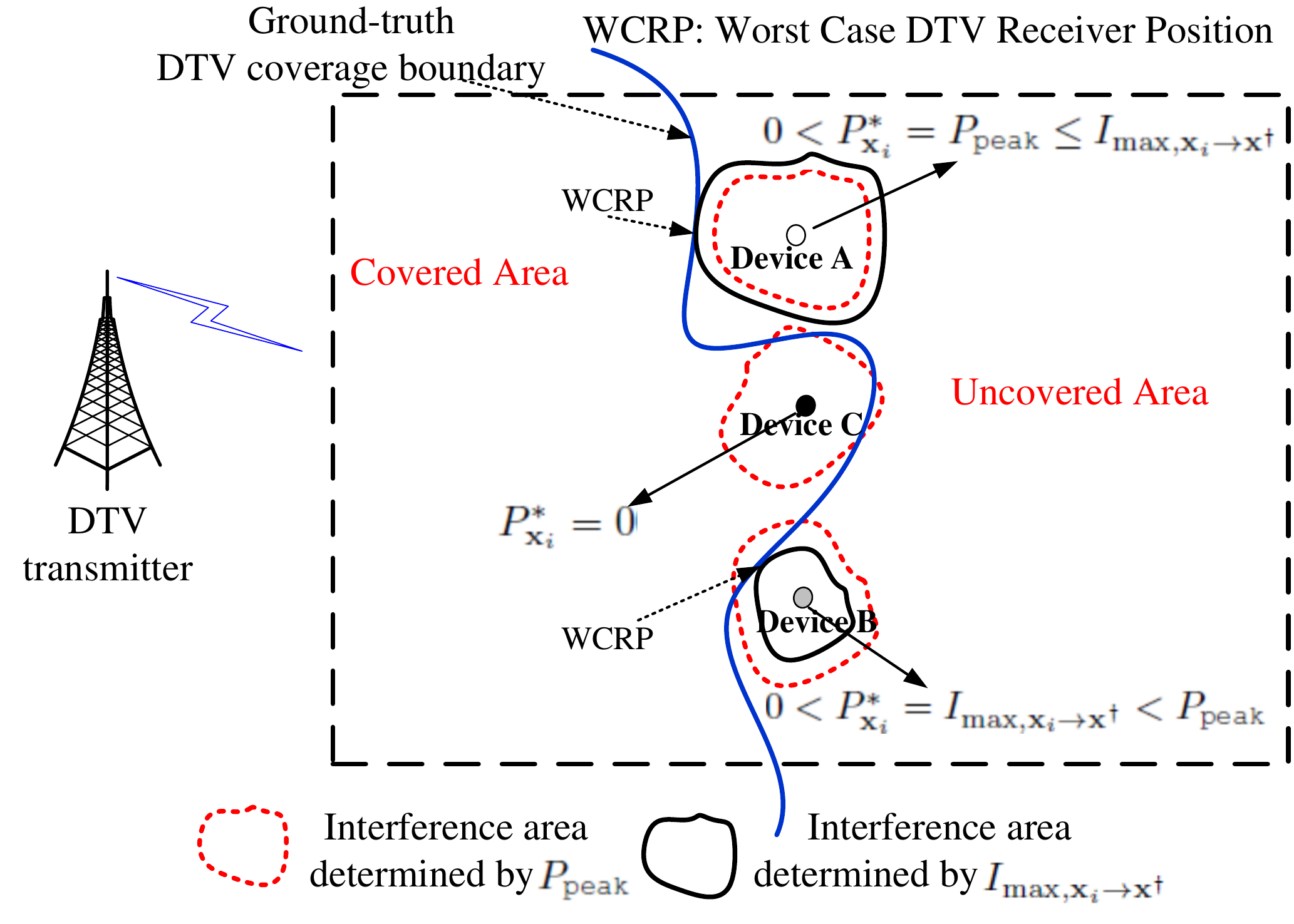}
\caption{Illustration for the optimal solution to problem $\textbf{\texttt{OP1}}$. The maximum permitted emission powers for device A, device B, and device C are ${P_{{\bf{x}}_i}^*}=P_{\texttt{peak}}$ (white space), ${P_{{\bf{x}}_i}^*} \in (0, P_{\texttt{peak}})$ (gray space), and ${P_{{\bf{x}}_i}^*}=0$ (black space), respectively.}
\label{Fig-WCRP}
\end{figure}

To facilitate the analysis, we introduce a notion named \emph{worst case DTV receiver position (WCRP)} for each unlicensed device. As shown in Fig. \ref{Fig-WCRP}, the WCRP for an unlicensed device is the location that lies on the boundary of the DTV coverage and perceives the strongest interference from that device. Now, we can rewrite (\ref{eq:Combination}) as
\begin{align}
\quad \quad \quad \quad P_{{\bf{x}}_i }  \le \min \{ {\rm{ }}P_{\texttt{peak} } ,I_{\max ,{\bf{x}}_i  \to {\bf{x}}^{\dag}}\},
\end{align}
where ${\bf{x}}^{\dag}$ denotes the WCRP for the device located at ${\bf{x}}_i$.

Intuitively, the optimal solutions to $\textbf{\texttt{OP1}}$ for devices located at various locations can be depicted as in Fig. \ref{Fig-WCRP}. It is noted that the optimization problem $\textbf{\texttt{OP1}}$ is quite difficult to tackle due to the following concerns:

(i) The boundary \emph{irregularity} of DTV coverage area makes it mathematically intractable to obtain an optimal solution. If the boundary is simplified to be circular as shown in Fig. \ref{fig_Visio-DTV-CBS-D2D}, positioning methods can be used to obtain an \emph{`optimal'} solution as in~\cite{BLMark_2009}. However, in practice, the ground-truth boundary is generally in arbitrary shapes and the optimal solution relies highly on the precise estimation of a local radio environment map~\cite{ZhaoYP_2006} for each unlicensed device, which is a nontrivial task for individual devices.

(ii) Strict interference constraint is required to protect licensed DTV receptions while \emph{limited} spectrum measurements are available for each unlicensed device. Collaborative sharing of measurement data among neighboring devices is a feasible way to improve the estimation of radio environment map for all devices~\cite{CGMap1_2011}. However, the energy and delay cost involved in the information exchange and the deployment cost of dedicated sensors may make it commercially unwelcome.

(iii) Spatial reuse opportunities for unlicensed devices at different locations are generally \emph{heterogeneous}. As pointed out in~\cite{TWC_mine}, at a given time, devices (even neighboring devices) at various locations may have different MPEPs (see the illustrative example in Fig. \ref{Fig-WCRP}), which makes a global solution infeasible.

These observations above motivate us to propose a technically feasible solution for the interesting but critical problem $\textbf{\texttt{OP1}}$. Specifically, Section IV provides an overview of the proposed approach. Section V details the algorithm designs and Section VI presents the performance evaluation.

\section{Overview of The Proposed Solution}

\subsection{Key Ideas}

One key insight is that, as an alternative to the existing propagation model-based geolocation database approach~\cite{Senseless_2012,Geo-Ding}, we can build a location-specific TVWS database by using active spectrum measurements with the aid of the already-deployed cellular infrastructure. Specifically, mobile devices equipped with programmable sensors can contribute massive \emph{location-aware} spectrum measurements and each cellular BS can collect and process big spectrum data to form a localized TVWS database, which in turn provides a spectrum \emph{service} for those mobile devices. Considering the fact that the spatial coverage of a DTV transmitter is relatively static, the spectrum measurements can be collected and updated in an \emph{offline} and \emph{asynchronous} manner. Moreover, with a large amount of collected spectrum data, it is possible to implement an extensive exploitation of TVWS by exploring intrinsic PHY-layer signal propagation characteristics of both distance-dominant path loss and location-dependent shadowing effect, which will undoubtedly provide more spatial reuse opportunities for small-scale D2D communications. In addition, recent advances in (big) data mining technologies make the proposed data-centric solution technically feasible.

\begin{figure}[!t]
\centering
\includegraphics[width=\linewidth]{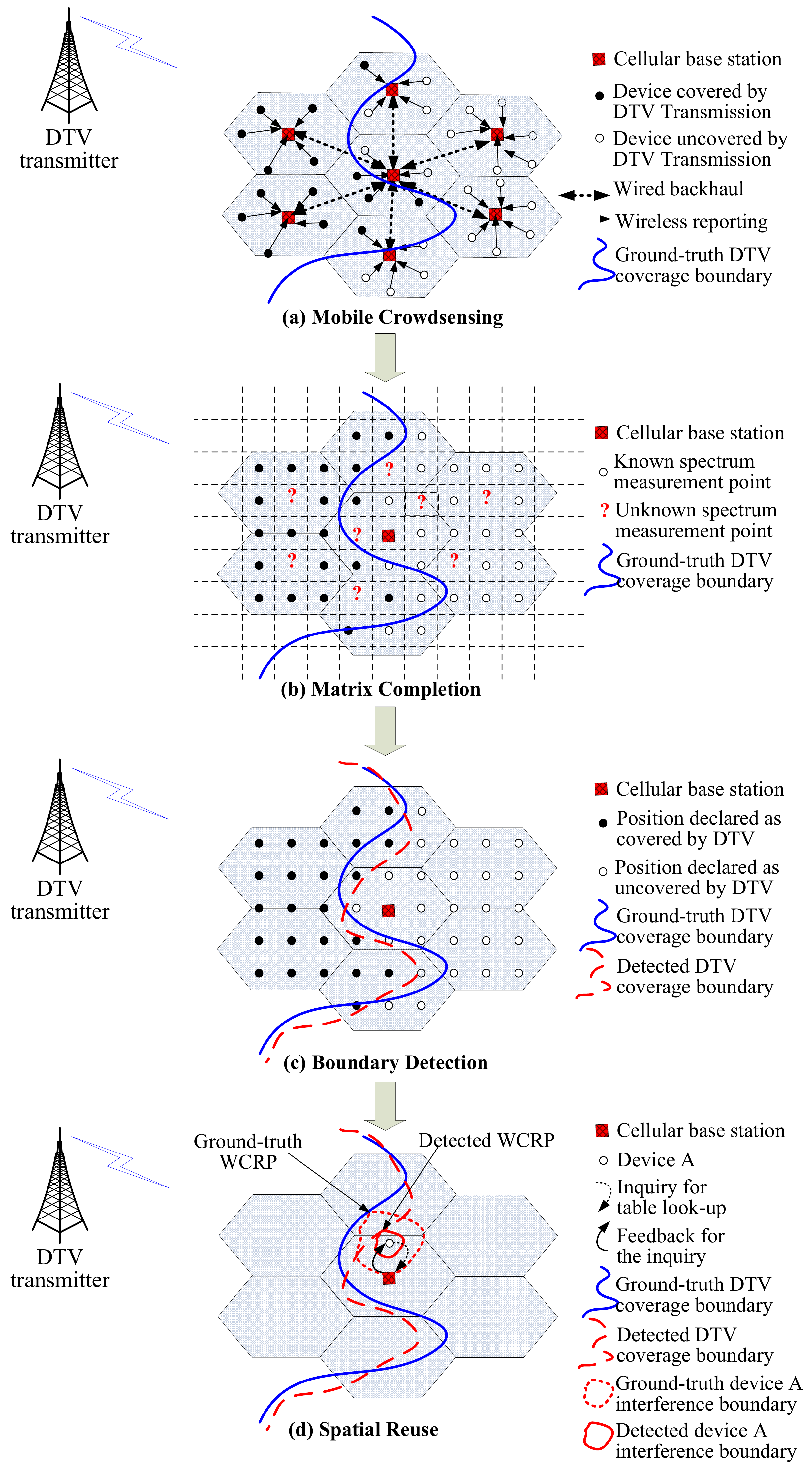}
\caption{The building blocks of the proposed solution.}
\label{Fig-Framework}
\end{figure}

\subsection{An Overview}
Fig. \ref{Fig-Framework} depicts an overview of the proposed solution, which can be implemented in the cellular system without requiring any modification to the existing infrastructure. Specifically, the proposed approach resides at each cellular BS and consists of the following sequential four building blocks:
\begin{itemize}
  \item \textbf{Mobile crowd sensing} that collects location-aware spectrum measurements from massive personal mobile devices with positioning capability.
  \item \textbf{Matrix completion} that recovers unknown spectrum data at locations that are lack of measurements based on some known spectrum data through effective spatial interpolation.
  \item \textbf{Boundary detection} that determines the irregular coverage boundary of the DTV transmitter by effectively mining the completed big spectrum data.
  \item \textbf{Spatial reuse} that enables the unlicensed D2D communication links to opportunistically share the same spectrum with the licensed DTV services.
\end{itemize}

\section{Algorithm Designs of The Proposed Solution}
In this section, we detail the algorithm design of each building block in the proposed solution.

\subsection{Mobile Crowd Sensing for Spectrum Measurements}
Spectrum measurement data is often obtained by deployed sensors with specialized equipments. Alternatively, in this paper, we introduce a novel information gathering technique, namely \emph{mobile crowd sensing (MCS)}~\cite{MCS2011}, to perform spectrum measurements with personal devices such as smartphones, tablets, and in-vehicle sensors. The key idea is that, with massive personal devices, each contribute a small amount of sensing data to ultimately obtain a sufficiently large dataset.

As shown in Fig. \ref{Fig-Framework}(a), in the proposed MCS-based spectrum measurement scheme, each mobile device uses a modern `\emph{mobile app}' to perform spectrum sensing and report the sensed DTV signal strength coupled with its current location information to its cellular BS through, e.g., a common control channel. Each cellular BS uses a local operational database to collect and store the spectrum measurement data.

Some key issues in MCS-based spectrum measurement scheme are discussed as follows.

\subsubsection{Trustiness of Contributions}
The first issue is how to ensure the quality of spectrum data from a large group of low-end personal devices that may be unreliable, untrustworthy, or even malicious. Specifically, the sensing report from a crowd spectrum sensor $m$ at location ${\bf{x}}$ can be expressed as:
\begin{equation}
y^{m}_{\bf{x}}  = {\underbrace{p_{{\bf{x}}_0  \to {\bf{x}}}^{m}+ v^{m}_{\bf{x}}}_{\text{Energy detector output},T^{m}_{\bf{x}}}} + {\underbrace {a^{m}_{\bf{x}}}_{\text{Abnormal data component}}}
\end{equation}
where $T^{m}_{\bf{x}}$ is the output of the energy detector including the received DTV signal component $p_{{\bf{x}}_0  \to {\bf{x}}}^{m}$ (in Watt) and the Gaussian noise component $v^{m}_{\bf{x}}$ with mean $N_0$ and variance $\sigma_{{v^{m}_{\bf{x}}}}^2=(p_{{\bf{x}}_0  \to {\bf{x}}}^{m} \cdot 1_{\{ \mathcal{H}_1\} }+ N_0)^2/N_{{\rm{sam}}}$. Notably, the abnormal data component is denoted as $a^{m}_{\bf{x}}$, which is zero if $y^{m}_{\bf{x}}$ is a normal data and nonzero if $y^{m}_{\bf{x}}$ is abnormal.

To effectively mitigate the uncertainty of sensing data from crowd spectrum sensors, in the previous work~\cite{TCOM-MINE-2014}, we have developed a data cleansing algorithm to robustly cleanse out the nonzero abnormal data component from the original corrupted sensing data, and moreover, in~\cite{SPMag_mine} we have designed a unsupervised clustering algorithm to distinguish spectrum attackers from reliable sensors based on dissimilarity analysis of historical reports, both of which can be applied to improve the quality of crowd spectrum data.

\subsubsection{Incentive to Contribute}
The second issue is how to motivate personal devices to participate in spectrum measurements and contribute the sensed data. While participating in a crowd sensing task, mobile devices consume their own resources such as battery and computing power, and expose themselves to potential privacy threats by sharing their sensed data with location information. However, we could consider a simple incentive mechanism that\emph{ devices which contribute spectrum data will receive a reward in terms of bandwidth and discounted or free communication minutes once TVWS is found available.} More advanced incentive mechanisms for general MCS systems, based on game theory and auction theory, can be found in~\cite{Incentive} and more general rules to motivate human cooperation can be found in~\cite{Nowak1,Nowak2}.

\subsubsection{Spectrum Data at Each Cellular BS}
\begin{figure}[!t]
\centering
\includegraphics[width=0.8\linewidth]{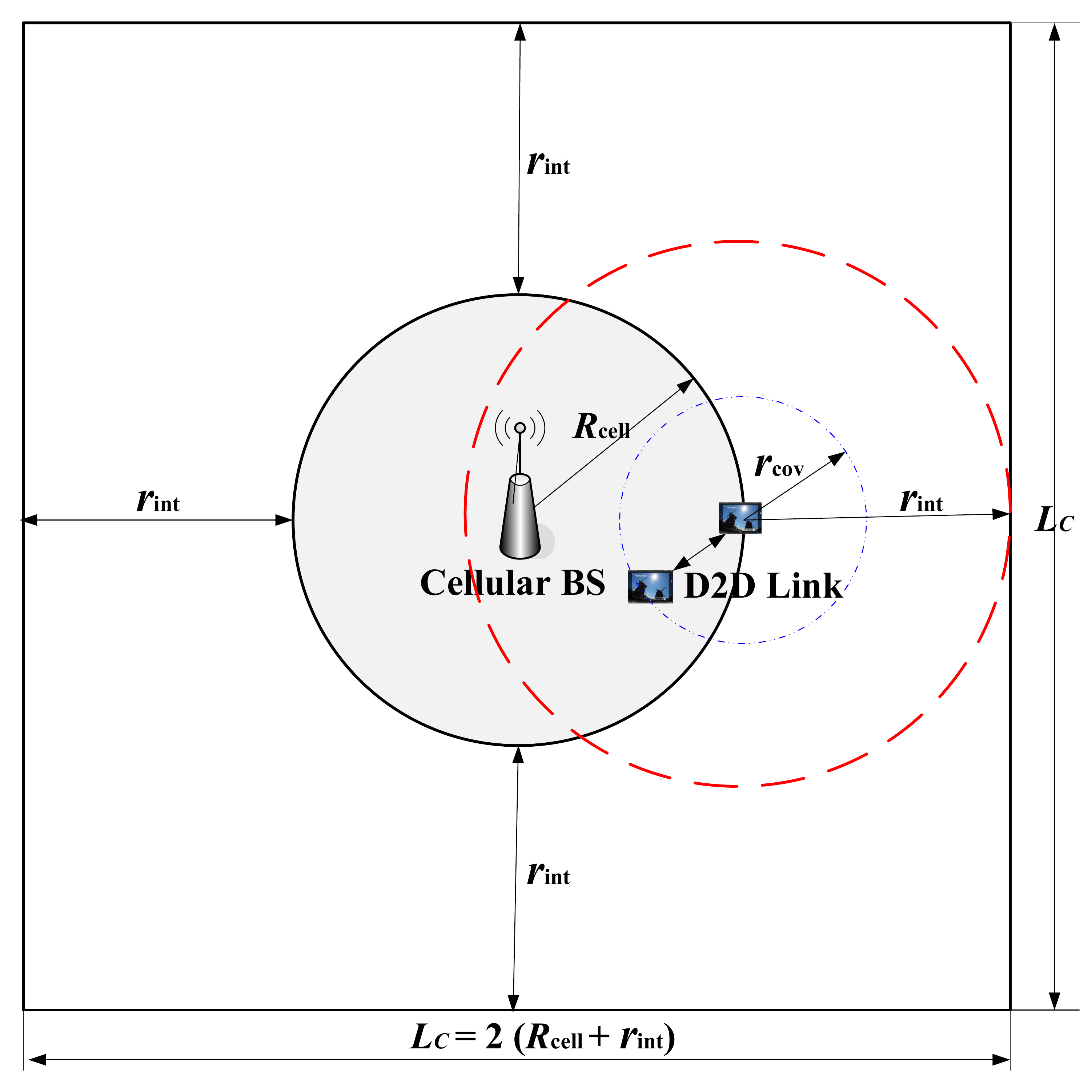}
\caption{Area of interest $\widetilde{\mathcal{A}}$ at each cellular BS. $R_{\rm{cell}}$ is the cellular radius, $r_{\rm{cov}}$ is the coverage range of a mobile device, and $r_{\rm{int}}$ denotes the interference range when the peak transmit power is used by the D2D link.}
\label{fig:CBS-Area}
\end{figure}

Another key issue is what type of spectrum data should be collected at each cellular BS. Briefly, the spectrum data at each cellular BS can be considered as multi-dimensional (i.e., time, frequency, and space) spectrum measurements (i.e., the received DTV signal power levels) from massive personal devices. Specifically, considering the spectrum data in time domain, for a given frequency band and a given location, spectrum measurements can be collected over multiple days in an offline and asynchronous manner. For the spectrum data in frequency domain, spectrum measurements can be from multiple TV channels. From the perspective of space domain, spectrum data at each cellular BS refers to spectrum measurements from an area of interest $\widetilde{\mathcal{A}}$, including both its cell (that can be directly collected from personal devices in its coverage area) and the neighboring cells (that can be obtained from the neighboring cellular BS's via backhaul). To facilitate the following processing, when we consider $\widetilde{\mathcal{A}}$ as a $L_C \times L_C$ square area as shown in Fig.~\ref{fig:CBS-Area}, the spatial range of $\widetilde{\mathcal{A}}$ is determined by the cellular radius $R_{\rm{cell}}$ and the \emph{worst-case interference range} $r_{\rm{int}}$ of a D2D link located at the edge of its cell, that is,
\begin{equation}
L_C=2(R_{\rm{cell}}+r_{\rm{int}}).
\end{equation}

\subsubsection{Uplink Overhead Analysis}
The use of mobile crowd sensing requires personal mobile devices to send their spectrum data to the cellular BS. This will result in additional uplink overhead in the cellular network for collecting spectrum measurements. Specifically, for a given cell, suppose that in a time period $T$, there are $N_{\rm{cell}}$ spectrum measurements collected by that cellular BS from $M_{\rm{cell}}$ mobile devices and each measurement is of $B$ bits, then the cell average uplink overhead is $\frac{N_{\rm{cell}}\times B}{T}$ bits/s, and the average uplink overhead for each device is $\frac{N_{\rm{cell}}\times B}{T\times M_{\rm{cell}}}$ bits/s. To decrease the uplink overhead, it is straightforward to increase $T$ and $M_{\rm{cell}}$, and decrease $N_{\rm{cell}}$ and $B$. Considering the fact that the spatial coverage of a DTV transmitter is relatively static, the spectrum measurements can be collected, from a variable number of mobile devices, in an \emph{offline} and \emph{asynchronous} manner over a relatively long time period $T$ (e.g., ranging from a few minutes to multiple days or even longer). Moreover, the effective spatial interpolator developed in the following subsection allows us to recover unknown spectrum data from a few known spectrum samples, which in essence decreases $N_{\rm{cell}}$.

\subsection{Matrix Completion for Unknown Measurements Recovery}
In MCS-based spectrum measurement scheme, the collected data at each cellular BS depends highly on individual human activities~\cite{human_mobility_pattern1,human_mobility_pattern2}. Consequently, there may be some locations or areas that are lack of (reliable) measurements. An effective spatial interpolator is vital for recovering unknown spectrum data from a few known samples. Most existing spatial interpolators can be represented as linear-weighted averages of sampled data and the weight designs are generally heuristic, such as the well-known inverse distance weighting~\cite{Statistics_1993} and Kriging weighting~\cite{Kriging_1986}. Many studies in various disciplines (see, e.g.,~\cite{SpatialInter_2008}) have shown that there is no simple answer regarding the choice of an appropriate weight design, because a method is ``best" only for specific situations, such as the spatial configuration of the data and the underlying assumptions.

To avoid the need of explicit weights design, in this paper, we formulate the problem of unknown measurements recovery as a matrix completion problem and propose to apply a powerful fixed point continuation algorithm (FPCA)~\cite{FPCA} to reconstruct the unknown elements from its known elements. As shown in Fig. \ref{Fig-Framework}(b), each cellular BS first divides the area of interest $\widetilde{\mathcal{A}}$ into a set of small square grids. In each grid, if the number of the collected (reliable) measurements (i.e., the sensed DTV signal strengths) is large enough (e.g., $>>$ 10), the mean of them can be taken as an estimation of the average received DTV signal strength in that grid. Otherwise, that grid will be set as `unknown'.

Furthermore, the spectrum dataset at each cellular BS can be written as a $p \times m$ matrix $\textbf{M}$, with the entry $\textbf{M}_{i,j}$ denoting the sensed average DTV signal strength in the $(i,j)$-th grid. Due to the reasons mentioned above, after collecting the spectrum measurements for a given time period, the cellular BS can only obtain a subset $\textbf{E} \subset [p] \times [m]$ of $\textbf{M}$'s entries. The partial observation of $\textbf{M}$ is defined as a $p \times m$ matrix $\textbf{M}^E$ with the entry as
\begin{align}
\textbf{M}^E_{i,j} = \begin{cases}
\textbf{M}_{i,j},  \quad \textrm{if}~(i,j)~\in~\textbf{E} \\
0, \quad \quad \textrm{otherwise}.
\end{cases}
\end{align}

We shall recover the unobserved elements of $\textbf{M}$ from $\textbf{M}^E$, which can be implemented by solving the following nuclear norm minimization£»
\begin{align}
\label{eq:NuclearNormOpt}
\min_{\textbf{M} \in \mathbb{R}^{p \times m}} \tau ||\textbf{M}||_{\ast}+\frac{1}{2}\sum_{(i,j)\in \textbf{E}} |\textbf{M}_{i,j}-\textbf{M}^E_{i,j}|^2,
\end{align}
where $||\textbf{M}||_{\ast}$ denotes the nuclear norm of matrix $\textbf{M}$ and $\tau$ is a scaling parameter that balances the first term and the second term in (\ref{eq:NuclearNormOpt}). For notational simplicity, a linear operator $\mathcal{P}$ is introduced to select the components $\textbf{E}$ out of a $p \times m$ matrix and form a vector such that $||\mathcal{P}(\textbf{M})-\textbf{M}^E||_2^2 = \sum_{(i,j)\in \textbf{E}} |\textbf{M}_{i,j}-\textbf{M}^E_{i,j}|^2$.

We adopt FPCA here to solve (\ref{eq:NuclearNormOpt}) for its outstanding performance in fast completion of large-scale matrices. Briefly, FPCA is based on the following fixed-point iteration~\cite{FPCA}:
\begin{align}
\label{eq:FPCA}
\begin{cases}
\textbf{Y}^k = \textbf{M}^k -\Delta \mathcal{P}^{*} (\mathcal{P}(\textbf{M}^k)-\textbf{M}^E) \\
\textbf{M}^{k+1} = S_{\tau\Delta}(\textbf{Y}^k),
\end{cases}
\end{align}
where $\Delta$ is a step size, $\mathcal{P}^{*}$ denotes the adjoint of $\mathcal{P}$, and $S_{\nu}(\cdot)$ is the matrix shrinkage operator (see Appendix A for details).

It is observed that the first step of (\ref{eq:FPCA}) is a gradient-descent applied to the second term in (\ref{eq:NuclearNormOpt}), which thus reduces the second term while generally increases the first term (i.e., the nuclear norm) in (\ref{eq:NuclearNormOpt}). In contrast, the second step of (\ref{eq:FPCA}) is to reduce the nuclear norm of $\textbf{Y}^k$. Iterations based on (\ref{eq:FPCA}) converge when the following stopping criterion is reached~\cite{FPCA}:
\begin{align}
\frac{||{\textbf{M}}^{k+1}-{\textbf{M}}^{k}||_F}{\max\{1,||{\textbf{M}}^{k}||_F\}} \le \beta,
\end{align}
where $||\cdot||_F$ is the Frobenius norm and $\beta$ is a small positive scalar (e.g., $10^{-6}$).

\begin{figure}[!t]
\centering
\includegraphics[width=\linewidth]{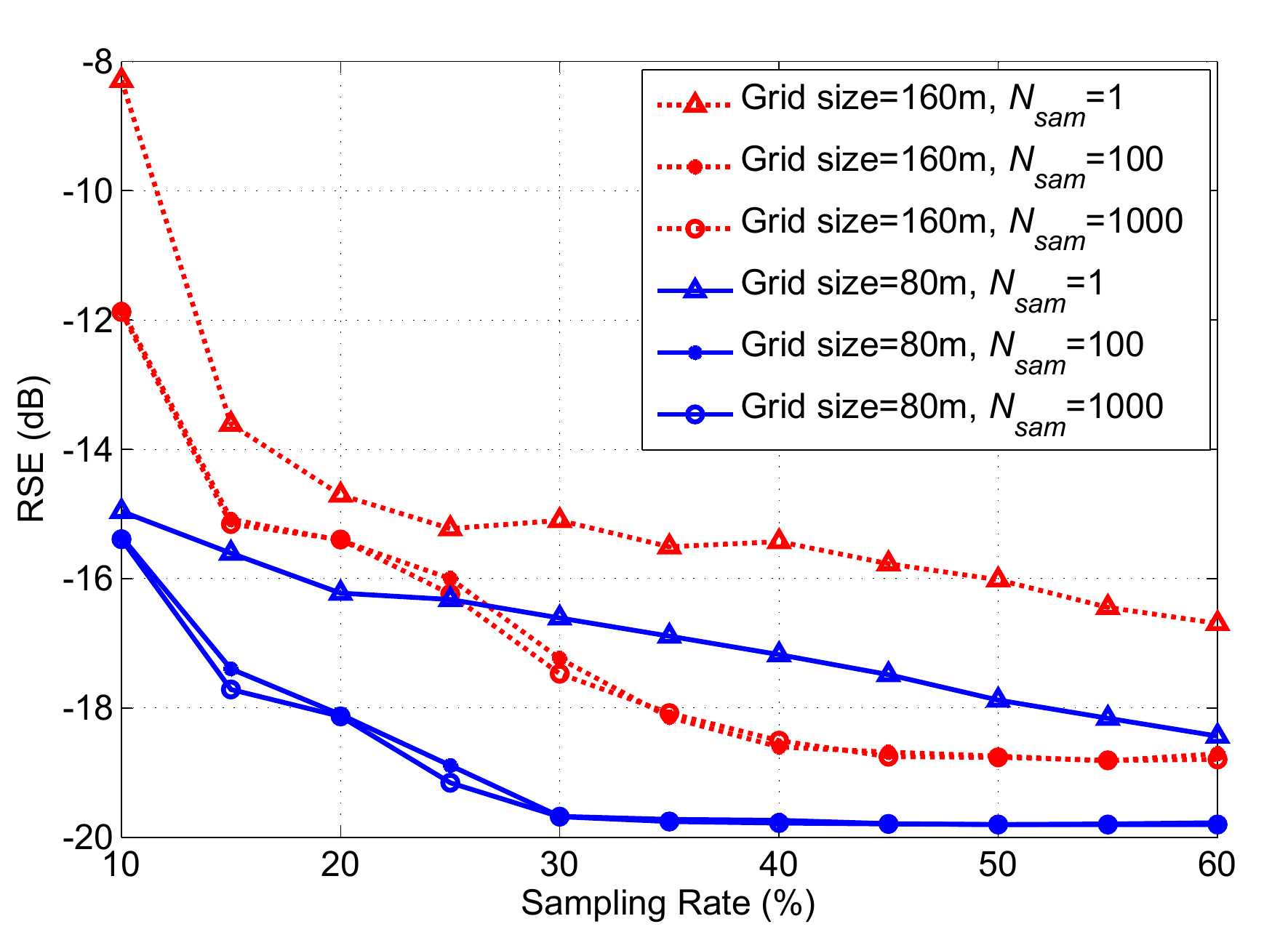}
\caption{The spectrum measurements recovery performance under various sampling rates. Here the sampling rate refers to the percentage of known elements in the spectrum data matrix ${\textbf{M}}$. The detailed parameter configuration of this simulation can be found in Section VI-A.}
\label{fig:MC-RSE}
\end{figure}

Fig.~\ref{fig:MC-RSE} shows the spectrum measurements recovery performance, in terms of recovery root square error (RSE), which is defined as
\begin{align}
RSE[dB] =10\log_{10}\frac{||\widetilde{\textbf{M}}-{\textbf{G}}||_2}{||{\textbf{G}}||_2},
\end{align}
where $\widetilde{\textbf{M}}$ denotes the recovered spectrum dataset and ${\textbf{G}}$ represents the ground-truth spectrum dataset with the entry ${\textbf{G}}_{i,j}$ as the mean received DTV signal strength at the center of the $(i,j)$-th grid. It is observed in Fig.~\ref{fig:MC-RSE} that:
\begin{itemize}
  \item The recovery error generally decreases with an increasing number of measurement samples in each grid (i.e., $N_{sam}$) as well as the sampling rate (i.e., the percentage of known elements in the total spectrum data matrix $\bf{M}$).
  \item Smaller grid size or higher spatial resolution yields better recovery performance.
  \item The recovery RSE approaches -20 dB when the spatial resolution is $80\times80$ m and the sampling rate is no smaller than 30 $\%$.
\end{itemize}

\emph{Computational complexity analysis}: In each iteration of (\ref{eq:FPCA}), a singular value decomposition (SVD) has to be computed to perform the matrix shrinkage operation, which has a high computational cost. Consequently, FPCA uses a rank-$r$ approximate SVD to replace the whole SVD, where $r$ is the estimated rank. Numerical experiments show that FPCA is very robust and not sensitive to the accuracy of the approximate SVDs. Moreover, a continuation strategy is adopted by FPCA to accelerate the convergence of (\ref{eq:FPCA}) and reduce the entire computation time, which solves a sequence of instances of (\ref{eq:NuclearNormOpt}), from easy to difficult, corresponding to a sequence of large to small values of $\tau$~\cite{FPCA}.

\subsection{Support Vector Machine for Coverage Boundary Detection}
Based on a complete matrix of spectrum data, in this section, we develop effective algorithms to reliably detect the DTV coverage boundary (see Fig. \ref{Fig-Framework}(c)), which is the key information for spatial reuse.

As mentioned in Section V-B, for a given cellular BS, the area of interest $\widetilde{\mathcal{A}}$ has been divided into a set of square grids. Specifically, the 2-D coordinates of the $(i,j)$-th grid center is denoted as $l_{i,j}:=(x_i,y_j)\in \widetilde{\mathcal{A}},i\in\{1,...,p\}, j\in\{1,...,m\}$ and the corresponding recovered average DTV signal strength is $\widetilde{\textbf{M}}_{i,j}$. Based on the energy detection, the cellular BS makes a binary test hypothesis for each grid as follows:
\begin{equation}
\label{eq:deltaP}
\widetilde{\textbf{M}}_{i,j} \underset{h(l_{i,j})=+1} {\overset{h(l_{i,j})=-1} \gtreqless} \bar{P},
\end{equation}
where $\bar{P}:=\bar{P}_{\min}-\delta_P$ is the detection threshold, with $\bar{P}_{\min}$ denoting the minimum average received signal power that a DTV receiver can reliably decode the desired signal, which has been defined in \textbf{Definition 1}, and $\delta_P$ being an offset parameter to compensate the imperfection of recovered spectrum data, which is positive for a conservative design to protect licensed DTV services and negative for an aggressive design to improve the unlicensed spectrum reuse. $h(l_{i,j})$ is a binary declaration on whether location $l_{i,j}$ is covered ($h(l_{i,j})=-1$) by the DTV transmission or not ($h(l_{i,j})=+1$).

Due to the hardware constraint of personal devices and radio channel randomness, the declarations are generally not error-free\footnote{As shown in Fig. \ref{Fig-Framework}(c), circles inside the ground-truth coverage and solids outside the ground-truth coverage are erroneous declarations.}. We, however, have no prior knowledge which declarations are correct. Based on the declarations with potential errors, the objective of coverage boundary detection is to find a function $f$ with the minimal detection errors, i.e.,
\begin{align}
\label{eq:op-boundary}
\mathop{\min}\limits_{f} \sum\nolimits_{l_{i,j} \in  \widetilde{\mathcal{A}}} {\bar h_f(l_{i,j}) \oplus h(l_{i,j})},
\end{align}
where $\bar h_f(l_{i,j})\in\{-1,+1\}$ denotes the coverage state at $l_{i,j}$ determined by the boundary function $f$ (i.e., if $f(l_{i,j})\ge1, \bar h_f(l_{i,j})=+1$; else if  $f(l_{i,j})\le -1, \bar h_f(l_{i,j})=-1$) and $\oplus$ is a binary operator defined as: if $\bar h_f(l_{i,j})=h(l_{i,j})$, $\bar h_f(l_{i,j}) \oplus h(l_{i,j})=0$; else if $\bar h_f(l_{i,j}) \ne h(l_{i,j})$, $\bar h_f(l_{i,j}) \oplus h(l_{i,j})=+1$.

To provide an efficient solution to (\ref{eq:op-boundary}), support vector machine (SVM)~\cite{Book_2000} serves as a promising theoretical tool. In the following we start with the formulation of the problem using simple linear SVM, and then extend to the design of nonlinear SVM classifiers by effectively kernelizing the linear SVM.

Specifically, linear SVM attempts to find a separating hyperplane (corresponding to a linear coverage boundary in this problem) $\langle {\bf{w}}, l \rangle+b=0, l\in\widetilde{\mathcal{A}}$ with the largest margin satisfying constraints:
\begin{align}
&\langle{\bf{w}},l_{i,j} \rangle +b \ge 0, ~{\rm{for}} ~h(l_{i,j})=+1;\nonumber \\
&\langle{\bf{w}},l_{i,j} \rangle +b \le 0, ~{\rm{for}} ~h(l_{i,j})=-1.
\end{align}
where ${\bf{w}}$ is a weight vector and $b$ is the intersect.

The optimal separating hyperplane can be derived by solving the following optimization problem~\cite{Book_2000}:
\begin{align}
\label{eq:opsvm}
&\mathop{\min}\limits_{{\bf{w}},b} ~\frac{1}{2}{\rm{ ||}}{\bf{w}}{\rm{||}}^{\rm{2}}\nonumber \\
&{\textrm{subject}}~{\textrm{ to }}~h(l_{i,j})(\langle{\bf{w}},l_{i,j}\rangle +b) \ge 1, l_{i,j}\in\widetilde{\mathcal{A}}.
\end{align}

To further consider the potential erroneous input (declarations), a regularization parameter $C>0$ is introduced to balance the tradeoff between the maximization of margin width and the penalty to the errors:
\begin{align}
\label{eq:op-svm2}
&\mathop{\min}\limits_{{\bf{w}},b,{\bf{\xi}}} ~\frac{1}{2}{\rm{ ||}}{\bf{w}}{\rm{||}}^{\rm{2}}  + C\sum\limits_{i = 1}^p \sum\limits_{j = 1}^m {\xi_{i,j} }\nonumber\\
&{\textrm{subject}}~{\textrm{ to }}~h(l_{i,j})(\langle{\bf{w}},l_{i,j}\rangle +b) \ge 1-\xi_{i,j}, l_{i,j}\in\widetilde{\mathcal{A}},
\end{align}
where ${\xi_{i,j} \ge 0}$ is a slack variable to reflect the impact of erroneous declarations. Furthermore, by introducing Lagrange multipliers $\alpha_{i,j}\ge0,i\in\{1,2,...,p\},j\in\{1,2,...,m\}$, the dual form of (\ref{eq:op-svm2}) can be expressed as
\begin{align}
\label{eq:oplinear}
&\mathop{\max} \limits_{\{\alpha_{i,j}\}}~ \sum_{i=1}^p \sum_{j=1}^m \alpha_{i,j} -\frac{1}{2}\sum_{i=1}^p \sum_{j=1}^m \sum_{k=1}^p \sum_{s=1}^m\alpha_{i,j} \alpha_{k,s} h(l_{i,j})h(l_{k,s}) \langle l_{i,j}, l_{k,s}\rangle\nonumber \\
&\textrm{subject~to}~\sum_{i=1}^p\sum_{j=1}^m \alpha_{i,j} h(l_{i,j})=0, 0 \le \alpha_{i,j} \le C.
\end{align}

The optimal solution $\{\alpha_{i,j}^{\star}\}$ to (\ref{eq:oplinear}) can be found by a quadratic programming solver~\cite{Book_2002}, and $\forall l\in \widetilde{\mathcal{A}}$, the decision function for the linear boundary can be expressed as:
\begin{align}
\label{eq:decision-function}
f(l)={\rm{sign}} (\sum_{i=1}^p\sum_{j=1}^m \alpha_{i,j}^{\star} h(l_{i,j})\langle l, l_{i,j}\rangle+b^{\star}),
\end{align}
where ${\rm{sign}}(\cdot)$ is the sign function and the threshold $b^{\star}$ can be obtained by averaging ${\tilde{b}_{k,s}}= h(l_{k,s})-\sum_{i=1}^p\sum_{j=1}^m \alpha_{i,j} h(l_{i,j})\langle l_{k,s}, l_{i,j}\rangle$ over all locations $l_{k,s}\in \widetilde{\mathcal{A}}$.

In practice, the ground-truth DTV coverage boundary is generally in \emph{nonlinear} and \emph{irregular} shape because of the signal attenuation resulted from obstructions such as hills or buildings~\cite{survey_PathLoss}. Thanks to \emph{kernel trick}~\cite{SPMag_mine}, we can derive more general nonlinear and irregular boundary functions by replacing the inner product operator in the linear SVM with an appropriate~\emph{kernel} function ${\rm{k}}(\cdot)$ as follows:
\begin{align}
\label{eq:kernel-trick}
\langle l_{i,j}, l_{k,s}\rangle \mapsto {\rm{k}}(l_{i,j},l_{k,s}),~\forall l_{i,j},l_{k,s}\in\widetilde{\mathcal{A}}.
\end{align}

Among others, the most widely used kernels include the (projective) polynomial kernel
\begin{align}
{\rm{k}}(l_{i,j},l_{k,s}) = (\langle l_{i,j}, l_{k,s}\rangle+c)^d,~c\ge0,d\in {\mathbb{N}}_{+};
\end{align}
and (radial basis function, RBF) Gaussian kernel
\begin{align}
{\rm{k}}(l_{i,j},l_{k,s}) = \exp(-|| l_{i,j}- l_{k,s}||_2^2/2\sigma^2),~\sigma>0;
\end{align}
where $c$, $d$, and $\sigma$ are kernel parameters, which are generally estimated from the training dataset~\cite{Book_2002,SPMag_mine}. These kernels implicitly map the data in original 2-D space onto a higher-dimensional feature space, where data are generally more separable. Consequently, the linear decision boundary function in (\ref{eq:decision-function}) can be extended to a nonlinear one as:
\begin{align}
\label{eq:irregular-boundary}
f(l)={\rm{sign}} (\sum_{i=1}^p \sum_{j=1}^m \alpha_{i,j}^{\star} h(l_{i,j}) {\rm{k}}(l, l_{i,j})+b^{\star}).
\end{align}

\begin{figure}[!t]
\centering
\includegraphics[width=\linewidth]{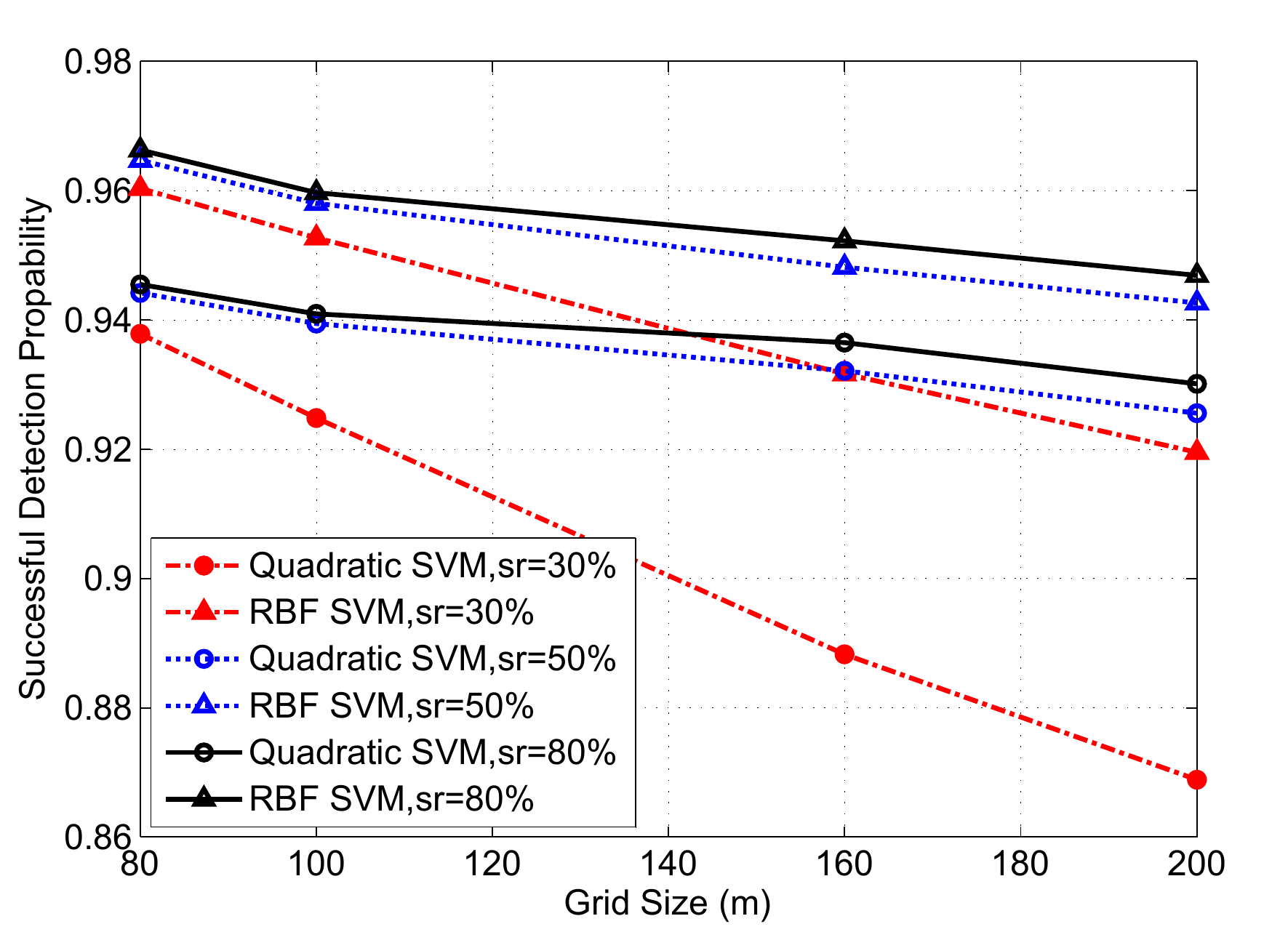}
\caption{The boundary detection performance in settings of various spatial resolutions. Each point is obtained by averaging 100 independent runs. In each run, uniform sampling is firstly implemented to obtain $\textbf{M}^E$ and matrix completion algorithm in Section V-B is then used to obtain $\widetilde{\textbf{M}}$. After that, SVM algorithms in Section V-C are used to derive the DTV coverage boundary.}
\label{fig:SVM-ErrorPro}
\end{figure}

From a statistical point of view, Fig.~\ref{fig:SVM-ErrorPro} presents the boundary detection performance of two representative SVM algorithms in terms of \emph{successful detection probability}, which denotes the probability that the grids inside the ground-truth DTV coverage area are detected as inside or the grids outside the coverage are detected as outside. It is shown in Fig.~\ref{fig:SVM-ErrorPro} that: i) Smaller grid size or higher spatial resolution yields better detection performance; ii) Higher sampling rate generally gives better performance; iii) RBF Gaussian SVM algorithm outperforms quadratic SVM for its superior capability of characterizing irregular boundary.

\emph{Computational complexity analysis}: The core of an SVM is a quadratic programming (QP) problem as shown in (\ref{eq:oplinear}) and the computational complexity of QP solvers is dataset dependent, scaling between ${\mathcal{O}}(N^2)$ and ${\mathcal{O}}(N^3)$~\cite{SVM-com}, where $N$ denotes the number of training samples.

\subsection{Opportunistic Spatial Reuse between a D2D Communication Link and the DTV Services}
As shown in Fig. \ref{Fig-Framework}(d), when the detected DTV coverage boundary is obtained, each cellular BS can compute the MPEP for any location inside its cell. Specifically, for any device located at ${\bf{x}}_i$ (inside the $(i,j)$-th grid), an opportunistic spatial reuse algorithm is proposed as follows:

\textbf{Case I:} If $h(l_{i,j})=-1$, we say grid $l_{i,j}$ is covered by the DTV or located in a \emph{black space} and thus no transmission is allowed to protect the potential DTV receptions, i.e., $P_{{\bf{x}}_i}^*=0$ (see, e.g., device C in Fig. \ref{Fig-WCRP});

\textbf{Case II:} $h(l_{i,j})=+1$ and there is no point of intersection between the detected DTV coverage boundary and the device's worst-case interference area that is determined by its peak transmit power $P_{\texttt{peak}}$, we say grid $l_{i,j}$ is located in a \emph{complete white space} and the device there can transmit in its peak transmit power ${P_{{\bf{x}}_i}^*=P_{\texttt{peak}}}$ (see, e.g., device A in Fig. \ref{Fig-WCRP});

\textbf{Case III:} $h(l_{i,j})=+1$ while there is a segment of the detected DTV coverage boundary covered by the device's worst-case interference area, we say grid $l_{i,j}$ is located in a \emph{gray space} and the device there can only transmit with the power ${P_{{\bf{x}}_i}^*} \in (0, P_{\texttt{peak}})$ (see, e.g., device B in Fig. \ref{Fig-WCRP}). Furthermore, ${P_{{\bf{x}}_i}^*}=I_{\max ,{\bf{x}}_i  \to {\bf{x}}^{\dag}}$. ${\bf{x}}^{\dag}$ denotes the WCRP for the device located at ${\bf{x}}_i$, which is the location that lies on the DTV coverage boundary and perceives the strongest interference from that device.

\emph{Remark 4:} In summary, based on the designed algorithms in Section IV, each cellular BS can compute and store the MPEPs for all locations inside its cell, which form a localized TVWS database. Any D2D communication that needs to transmit in an unlicensed TV spectrum band can submit an inquiry associated with its current location to its cellular BS. After receiving the inquiry request, the cellular BS makes a table look-up operation in its database and sends back the information of the MPEP that device can use without disrupting the normal operation of the licensed DTV services.

\section{Performance Evaluation}

\subsection{Simulation Setup}
We consider a wireless environment that a cellular network of small-scale mobile devices opportunistically and spatially reuse a TV channel licensed to a large-scale DTV system. Table I lists the system parameters used in the following simulations. The parameters of the DTV system and D2D communications are mainly based on the specifications in~\cite{Inter_Ana2009} and~\cite{D2D_2009,TSP-WC-Book}, respectively. The grid size, sampling rate, and number of measurements in each grid are based on the results obtained in Section V-B and V-C.

\begin{table}[!t]
\renewcommand{\arraystretch}{1}
\caption{System parameters used in simulations}
\label{table_example}
\centering
\begin{tabular}{c|c|c}
\hline
\bfseries Parameter & \bfseries Value & \bfseries Comment\\
\hline
\hline
$B$ & 6 MHz &Bandwidth of a TV channel \\
\hline
$f$ & 615 MHz &Center frequency of the TV channel \\
\hline
$P_t^{DTV}$ & 90 dBm &Transmission power of a DTV transmitter \\
\hline
$(x_0,y_0)$ & (0,0) &2-D coordinates of the DTV transmitter \\
\hline
$N_0$ & -95.2 dBm &Noise power \\
\hline
$d_p$ & 134.2 km &Radius of DTV protection region \\
\hline
$\alpha^{DTV}$ & 4 &Path loss exponent of DTV transmission\\
\hline
$\alpha^{D2D}$ & 2.5 &Path loss exponent of D2D transmission\\
\hline
$\sigma$ & 5.5 dB &Shadowing dB-spread \\
\hline
$P_{\min}$ & -92.2 dBm &Minimum power to decode the desired signal\\
\hline
$I_{\max}$ & -98.2 dBm &Interference power tolerance threshold\\
\hline
$\nu_{\rm{cov}}$ & 0.9 &Location coverage probability threshold \\
\hline
$\nu_{\rm{int}}$ & 0.1 &Location interference probability threshold \\
\hline
$\widetilde{\mathcal{A}}$ & 8 km $\times$ 8 km  &Area of interest at each cellular BS \\
\hline
$L_g$ & 80 m &Grid size\\
\hline
$sr$ & 50$\%$ &Sampling rate \\
\hline
$N_{sam}$ & 100 &Number of measurements in each grid \\
\hline
$R_{\rm{cell}}$ & 2 km  &Cellular radius of each cell \\
\hline
$P_{\rm{peak}}$ & -10 dBm  &Peak transmit power of a personal device \\
\hline
$r_{\rm{int}}$ & 2 km  &Worst-case interference range of a device \\
\hline
\end{tabular}
\end{table}

\begin{figure*}[!t]
\centering
\subfloat[Ground truth coverage]{
\label{fig:stacksub:a}
\includegraphics[width=0.25\linewidth]{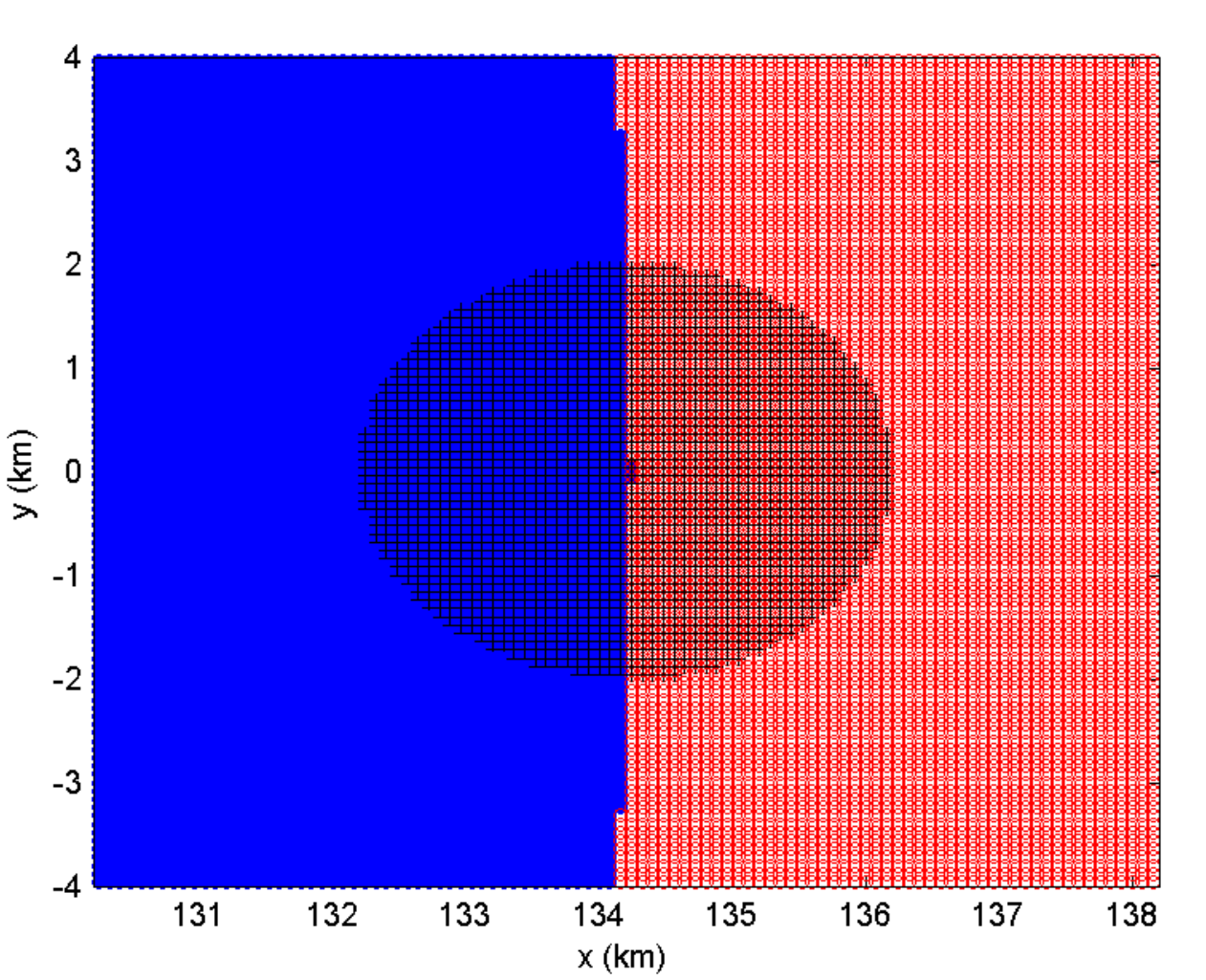}}
\subfloat[After random sampling]{
\label{fig:stacksub:b}
\includegraphics[width=0.25\linewidth]{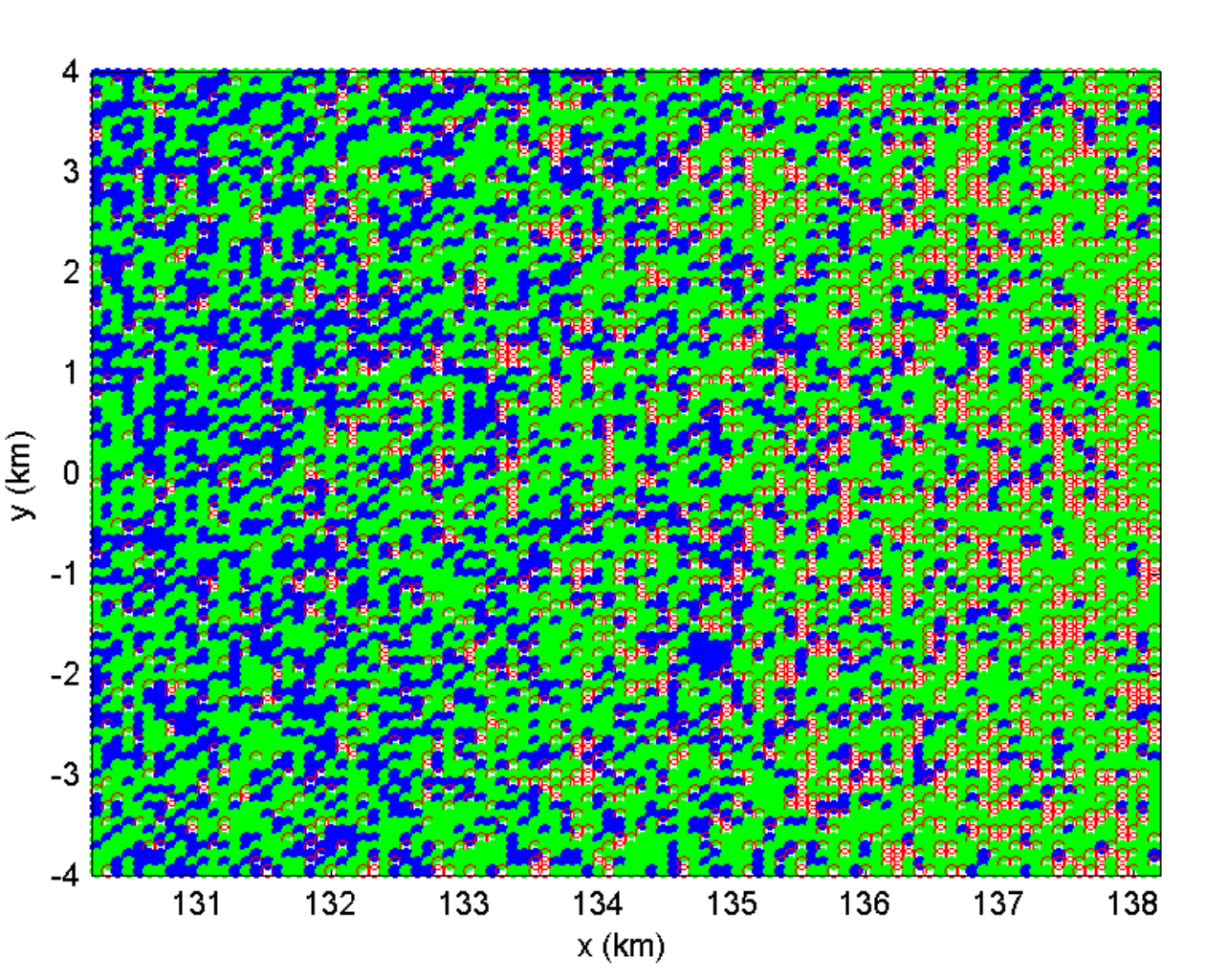}}
\subfloat[After matrix completion]{
\label{fig:stacksub:c}
\includegraphics[width=0.25\linewidth]{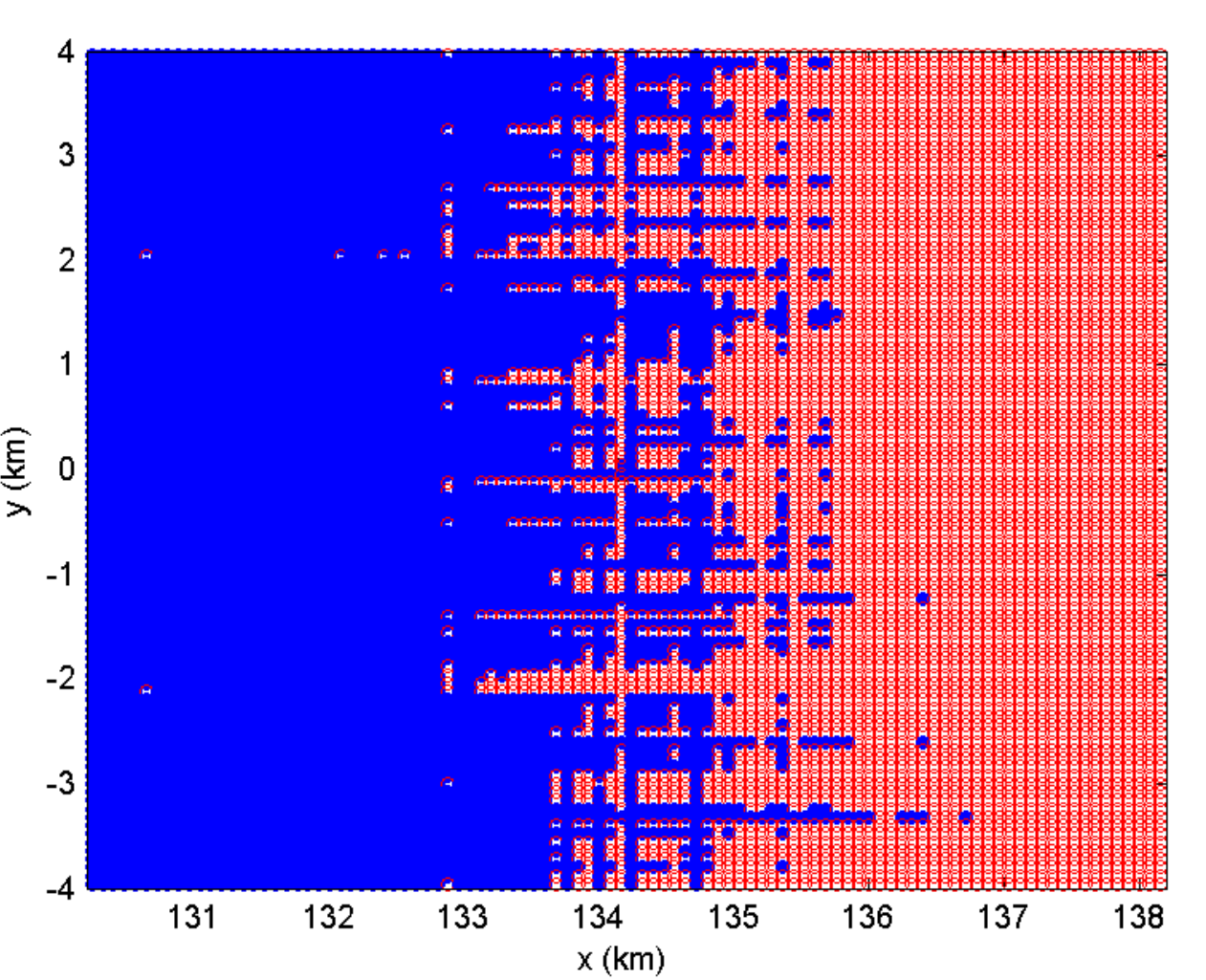}}
\subfloat[After SVM]{
\label{fig:stacksub:d}
\includegraphics[width=0.25\linewidth]{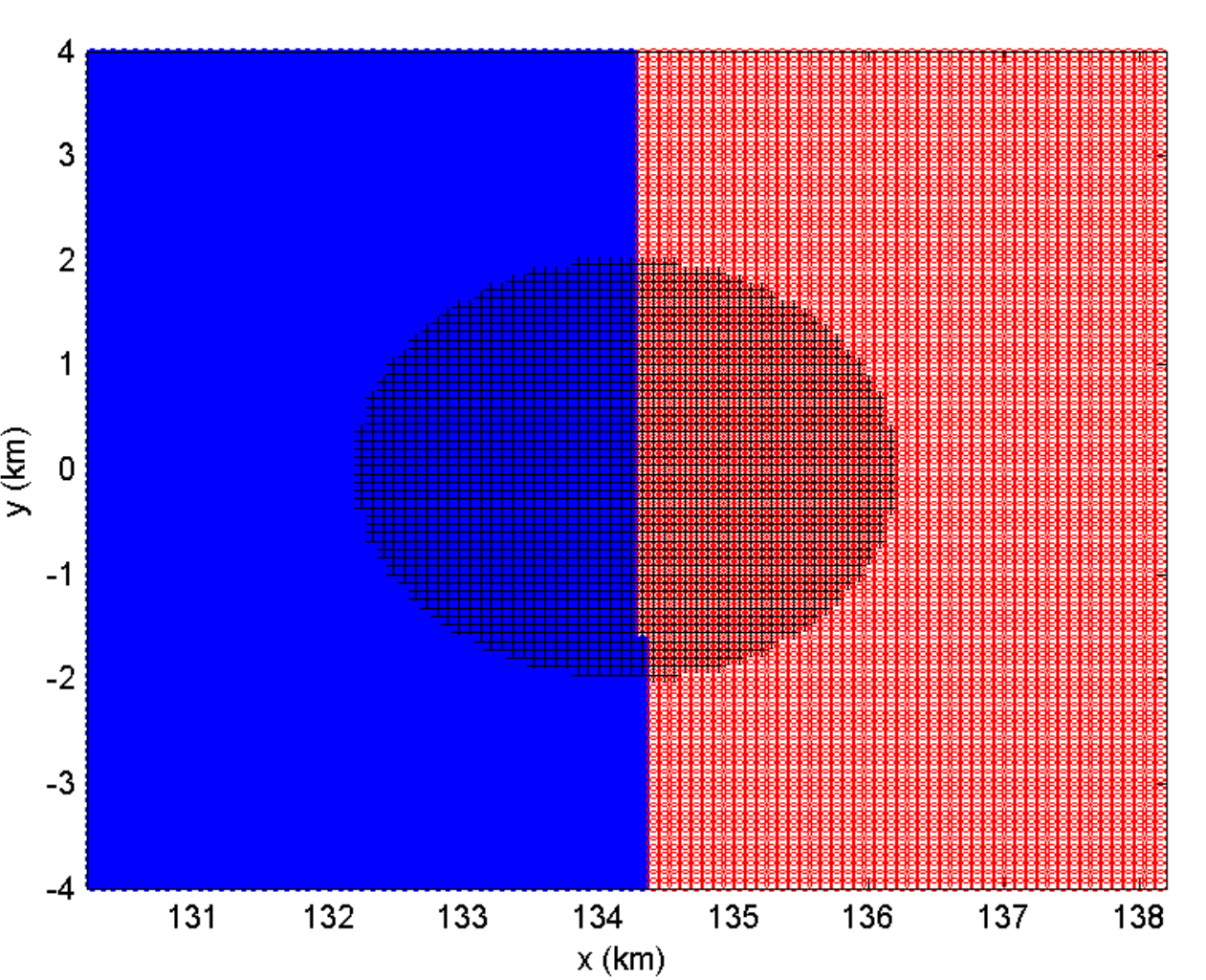}}
\caption{Simulation results of the proposed approach under \textbf{Scenario I} (Part I). The shaded disc area in subfigure(a) and subfigure(d) (around the centered cellular BS) represents the cell range of the cellular BS.}
\label{fig_Sim_Case_I}
\end{figure*}

For each cellular BS, the area of interest $\widetilde{\mathcal{A}}$ can be located i) \emph{completely inside} the DTV coverage, or ii) \emph{completely outside} the DTV coverage, or iii) \emph{partially inside partially outside} the DTV coverage. Technically, the third case is much more challenging to enable the co-channel deployment of a DTV system and D2D communications, since the interference management in this case is nontrivial\footnote{It is noted that interference management for D2D communications in TV spectrum is more challenging than in cellular spectrum, since the explicit coordination/cooperation from the licensed DTV users is unavailable.}. Consequently, in the following simulations, we will focus on two representative scenarios of the third case:

\textbf{Scenario I:} The cellular BS is located at the edge of the DTV protection region, which is also named as the noise-limited coverage contour in~\cite{20dB_2006,Inter_Ana2009}. In this scenario, any D2D link located inside the DTV protection region is not allowed to transmit, while D2D links located outside can be allowed to transmit without causing harmful interference to the potential DTV receptions inside the DTV protection region.

\textbf{Scenario II:} The cellular BS is located inside the DTV protection region, however, a portion of its cell falls into a shadowing zone resulted from large obstacles such as mountains or a group of buildings. In this scenario, DTV receptions cannot work in the shadowing zone, which thus forms white space for small-scale D2D communications. Any D2D link located inside the shadowing zone should be allowed to transmit without introducing harmful interference to the DTV receptions outside the shadowing zone.

\begin{figure*}[!t]
\centering
\subfloat[$\text{Distribution of biases of MPEPs}$.]{\includegraphics[width=0.5\linewidth]{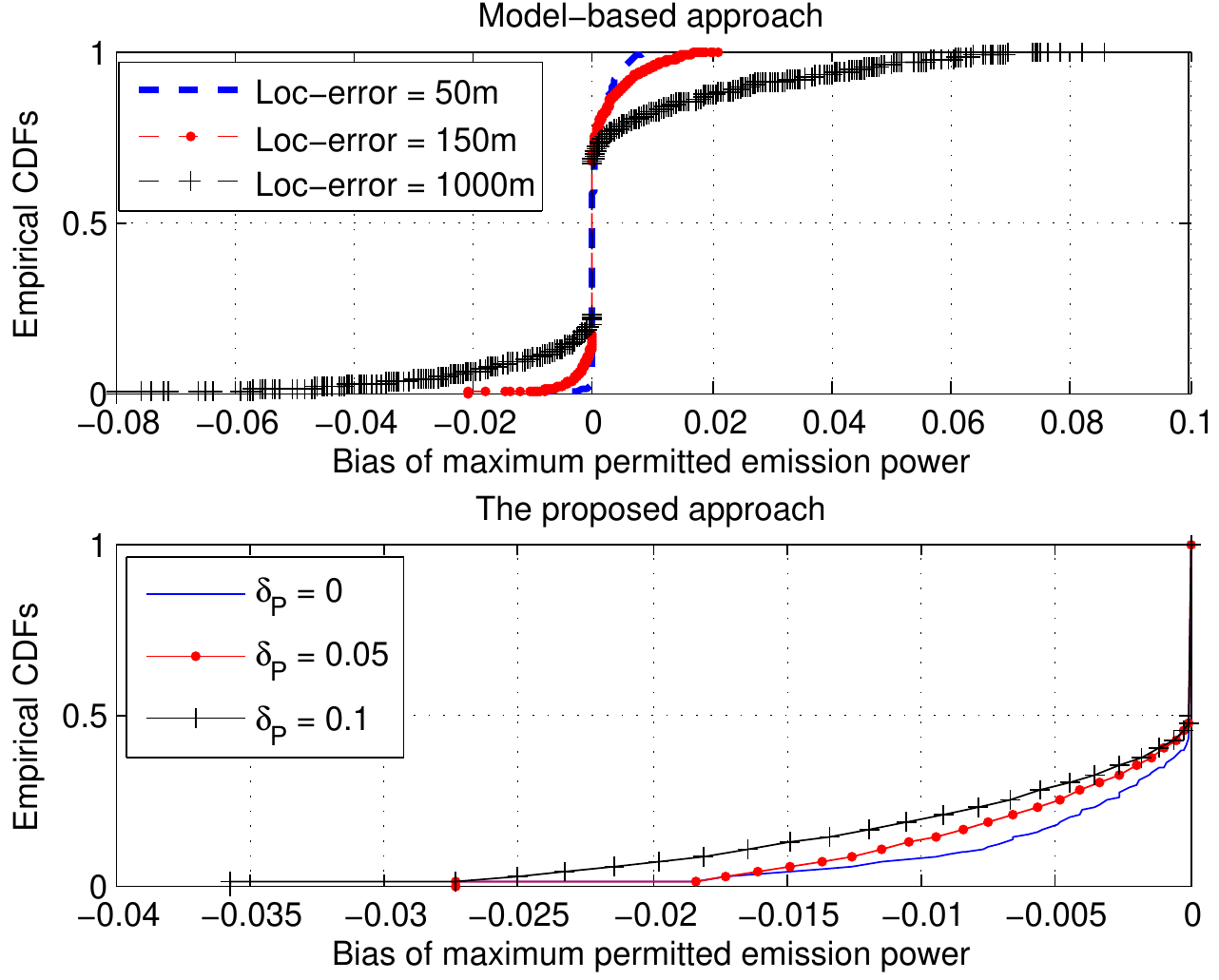}
\label{fig:stacksub:a}}
\subfloat[$\text{Distribution of biases of IPs}$.]{\includegraphics[width=0.5\linewidth]{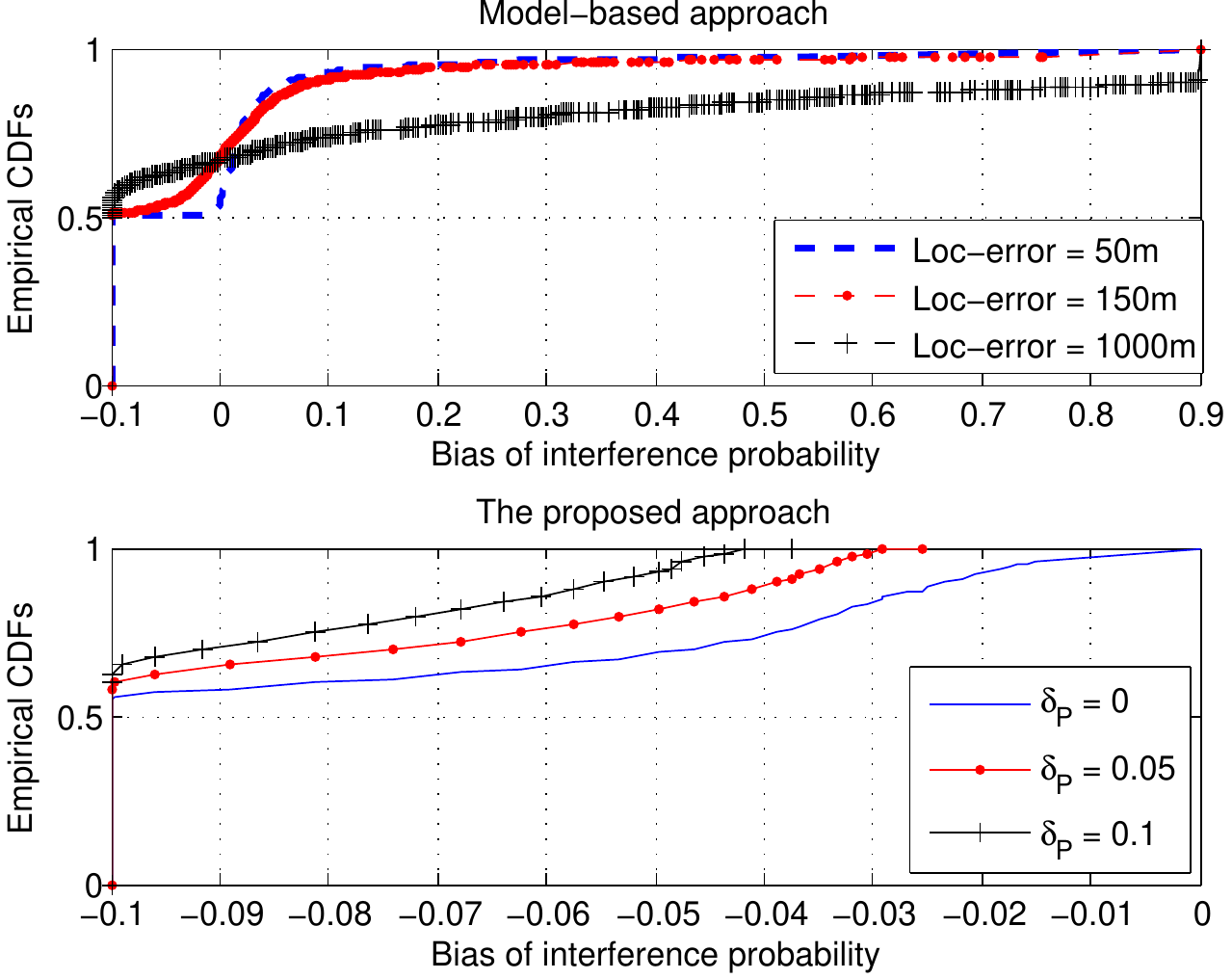}
\label{fig:stacksub:a}}
\caption{Simulation results under \textbf{Scenario I} (Part II). Considering all the locations inside the cell (i.e., the shaded disc area in Fig.~\ref{fig_Sim_Case_I}(a)) of the centralized cellular BS, from the statistical view, the left subfigure shows the distribution of the derived MPEP minus the corresponding ground-truth MPEPs, and the right subfigure shows the distribution of the derived interference probabilities minus the interference probability threshold $\nu_{\rm{int}}=0.1$.}
\label{fig_CaseI_MIFTP_IP}
\end{figure*}

\begin{figure*}[!t]
\centering
\subfloat[Ground truth coverage]{
\label{fig:stacksub:a}
\includegraphics[width=0.25\linewidth]{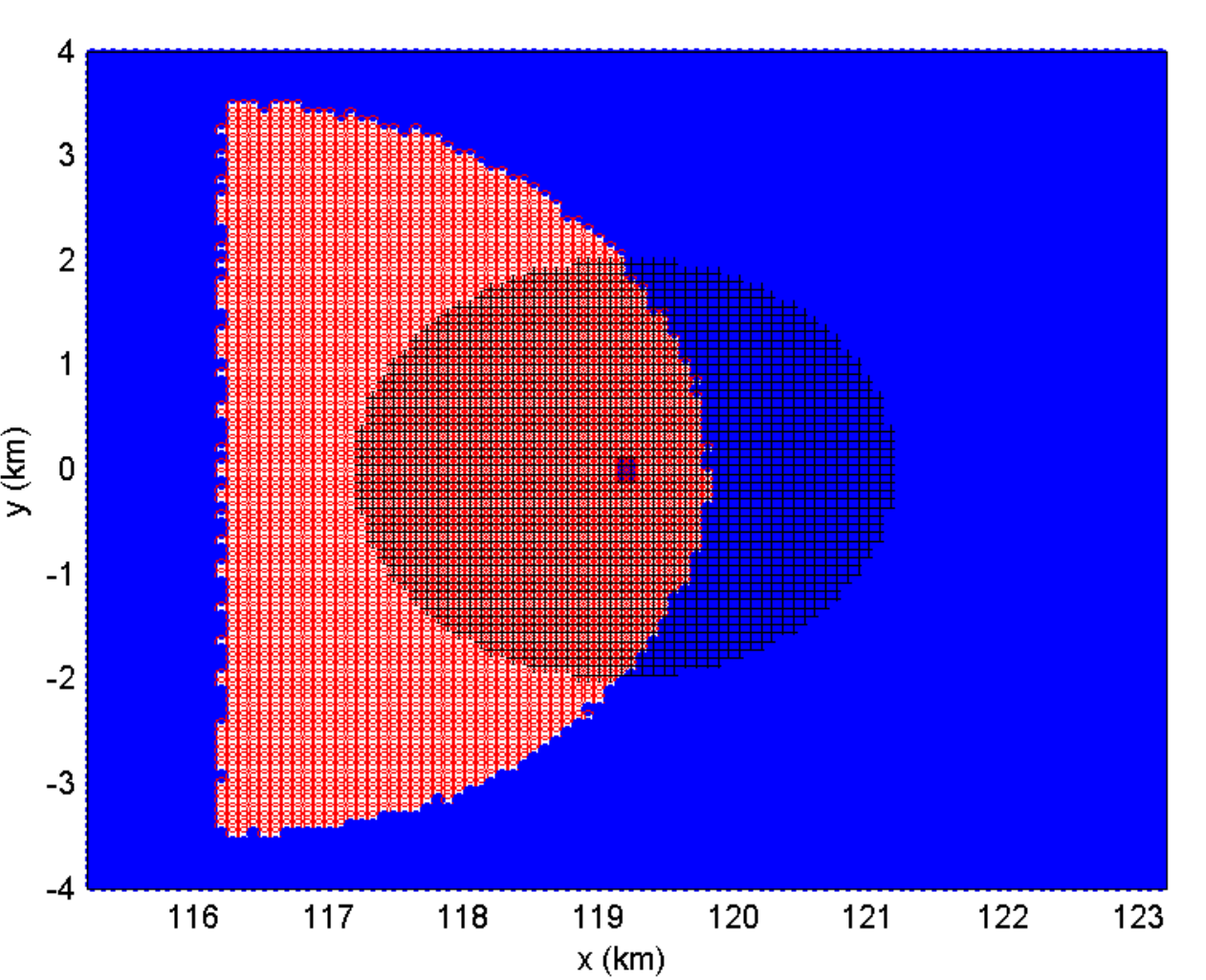}}
\subfloat[After random sampling]{
\label{fig:stacksub:b}
\includegraphics[width=0.25\linewidth]{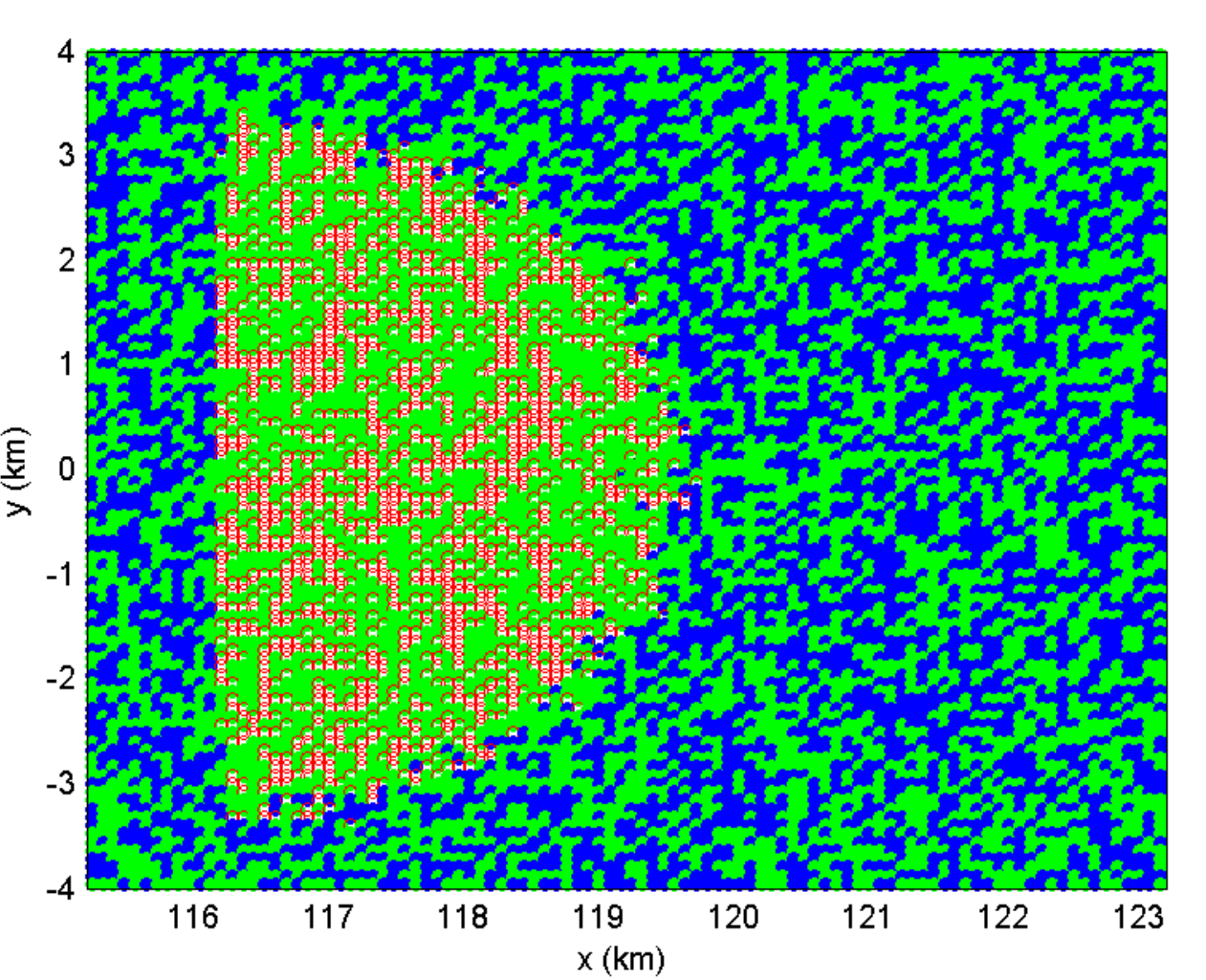}}
\subfloat[After matrix completion]{
\label{fig:stacksub:c}
\includegraphics[width=0.25\linewidth]{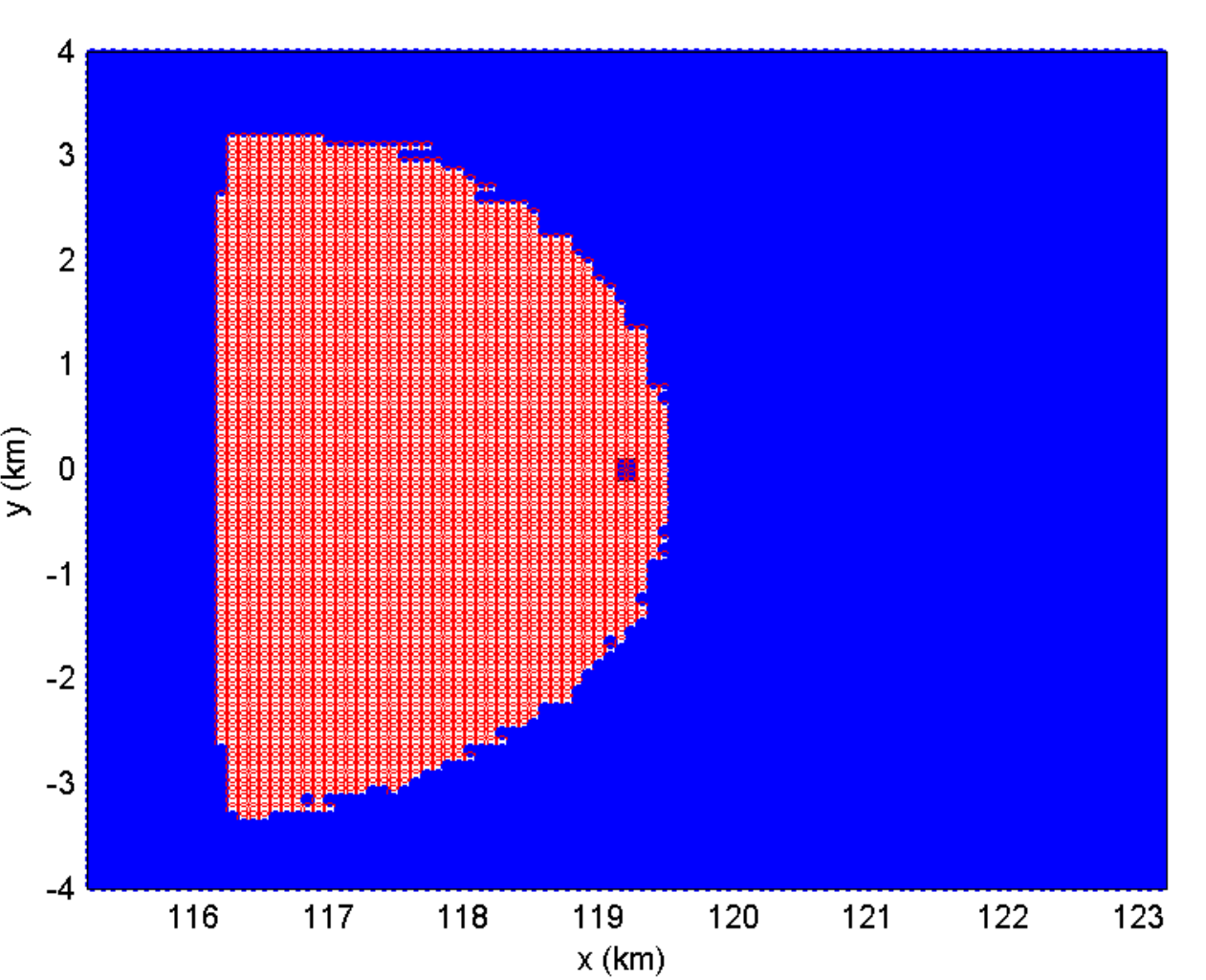}}
\subfloat[After SVM]{
\label{fig:stacksub:d}
\includegraphics[width=0.25\linewidth]{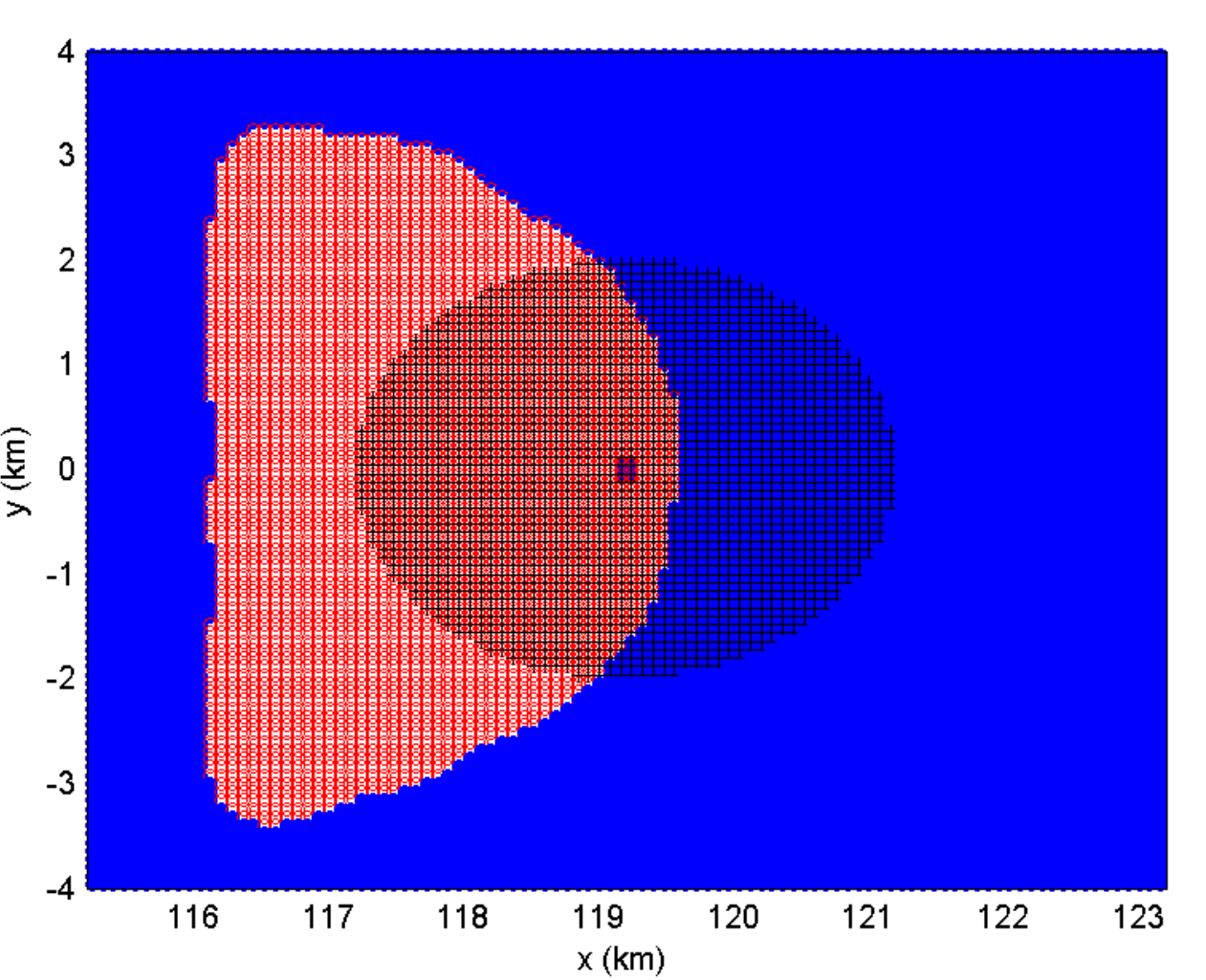}}
\caption{Simulation results of the proposed approach under \textbf{Scenario II} (Part I). The shaded disc area in subfigure(a) and subfigure(d) (around the centered cellular BS) represents the cell range of the cellular BS.}
\label{fig_Sim_Case_II}
\end{figure*}

\subsection{Simulation Results under Scenario I}
Under the representative simulation Scenario I, Fig.~\ref{fig_Sim_Case_I} presents the results to show the role of each building block algorithm of the proposed approach. In this simulation, we first generate the ground-truth DTV coverage (see Fig.~\ref{fig_Sim_Case_I}(a)) as the baseline reference. A cellular BS is assumed to be at the center of the 8 km $\times$ 8 km area $\widetilde{\mathcal{A}}$, which is further divided into 100 $\times$ 100 small grids. It appears that the ground-truth DTV coverage boundary (separating the blue and red areas) is almost linear since the cell (the shaded disc area) in Scenario I is located at the edge of the DTV coverage and the range of the DTV coverage is much larger than the range of the cell.

Then, mobile crowd sensing is used at the cellular BS to collect spectrum samples and only 50$\%$ locations are considered to be sampled in Fig.~\ref{fig_Sim_Case_I}(b). Intuitively, the DTV coverage is quite difficult to be recognized from the highly uncompleted and noisy spectrum data. To tackle this challenge, the matrix completion algorithm developed in Section V-B is used to recover the unknown measurements (see Fig.~\ref{fig_Sim_Case_I}(c)). It is shown that the recovered DTV coverage in Fig.~\ref{fig_Sim_Case_I}(c) becomes more recognizable than that in Fig.~\ref{fig_Sim_Case_I}(b), however, the boundary separating the recovered blue and red areas is still quite blur. Therefore, the SVM algorithm developed in Section V-C is further used to derive the DTV coverage boundary (see Fig.~\ref{fig_Sim_Case_I}(d)). Through comparing Fig.~\ref{fig_Sim_Case_I}(d) and Fig.~\ref{fig_Sim_Case_I}(a), it seems the proposed approach performs quite well.

Furthermore, Fig.~\ref{fig_CaseI_MIFTP_IP} presents additional results to show the effectiveness of the proposed approach over the traditional model-based approach~\cite{BLMark_2009,Senseless_2012}. Specifically, considering all the locations inside the cell (i.e., the shaded disc area in Fig.~\ref{fig_Sim_Case_I}(a)) of the centralized cellular BS, from the
statistical view in terms of empirical cumulative distribution functions (CDFs), Fig.~\ref{fig_CaseI_MIFTP_IP}(a) shows the distribution of the derived MPEPs at various locations minus the corresponding ground-truth MPEPs, and Fig. 8(b) shows the distribution of the derived interference probabilities (IPs) at various locations minus the IP threshold $\nu_{\rm{int}}=0.1$.

It is shown in Fig.~\ref{fig_CaseI_MIFTP_IP} that:
\begin{itemize}
  \item In the traditional model-based approach, the biases of MPEPs range from -0.08 to 0.1 and the biases of IPs range from -0.1 to 0.9, and the distributions of both of the biases are closely related to the specific localization error, which is a key parameter to limit the performance of the model-based approach. According to~\cite{Senseless_2012}, the simulation setup of the maximum localization error is 50m for GPS, 150m for Wi-Fi, and 1000m for GSM-based localization, respectively.
  \item Relatively, the proposed approach has much smaller ranges for both of the biases of MPEPs and the biases of IPs, and the maximum of both of the biases can be smaller than 0 by adjusting the offset parameter $\delta_P$ (defined in Eq.(\ref{eq:deltaP})), which means that the proposed approach can successfully enable D2D communications in TVWS while satisfying the interference constraint from the licensed DTV services. Moreover, as mentioned in Section V-III, $\delta_P$ is a parameter to compensate the imperfection of recovered spectrum data. Generally, a larger $\delta_P$ corresponds to a more conservative design, which means the derived MPEPs are much smaller than the ground truth ones and the corresponding IPs are much smaller than the threshold $\nu_{\rm{int}}=0.1$.
\end{itemize}

In brief, compared with the traditional model-based approach, the advantages of the proposed approach are mainly twofold: i) improved spatial reuse between large-scale DTV services and small-scale D2D links can be obtained; ii) reduced interference to the DTV receptions near the edge of the ground-truth coverage area could be achieved.

\subsection{Simulation Results under Scenario II}
\begin{figure*}[!t]
\centering
\subfloat[$\text{Distribution of biases of MPEP levels}$.]{\includegraphics[width=0.5\linewidth]{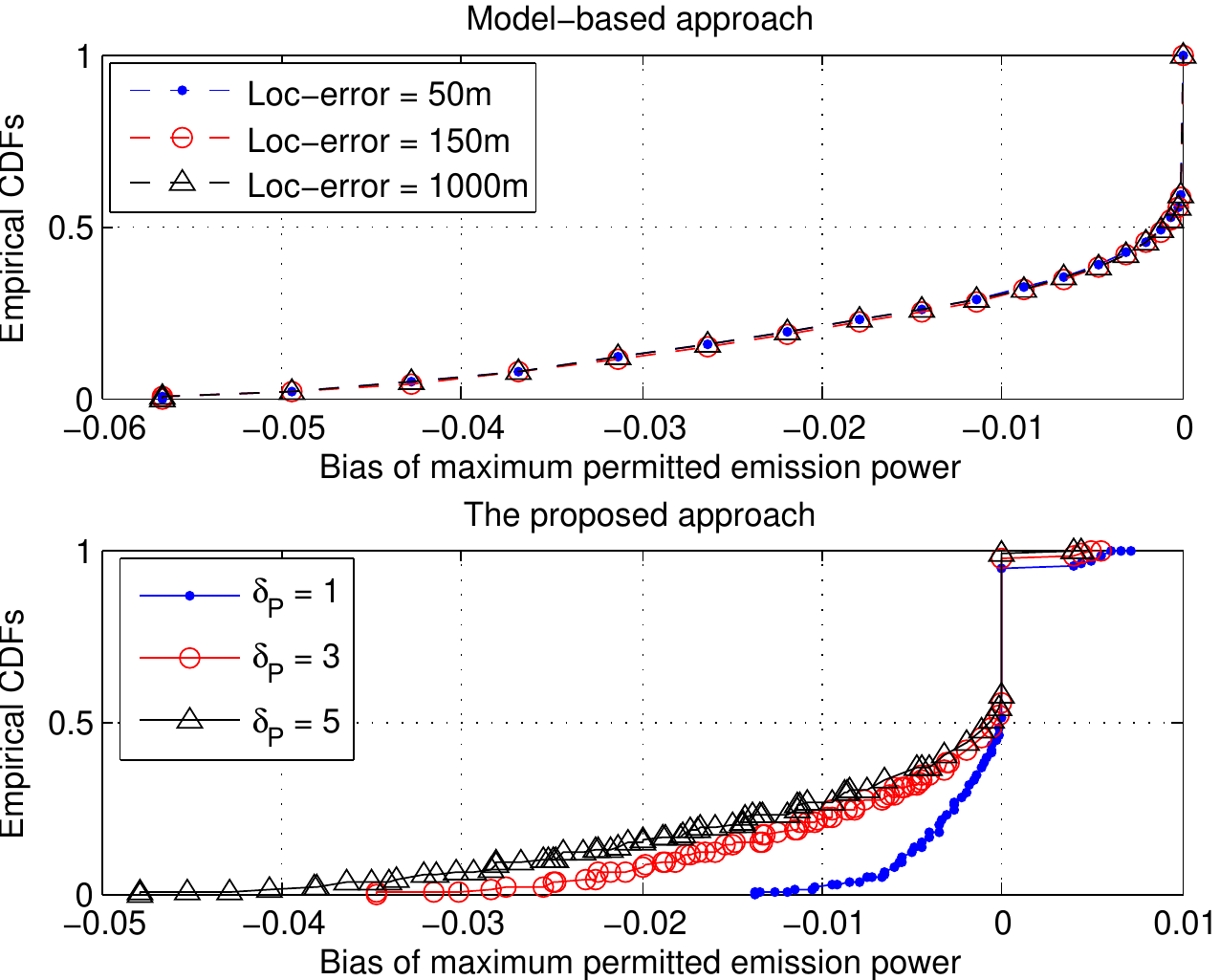}
\label{fig:stacksub:a}}
\subfloat[$\text{Distribution of biases of IPs}$.]{\includegraphics[width=0.5\linewidth]{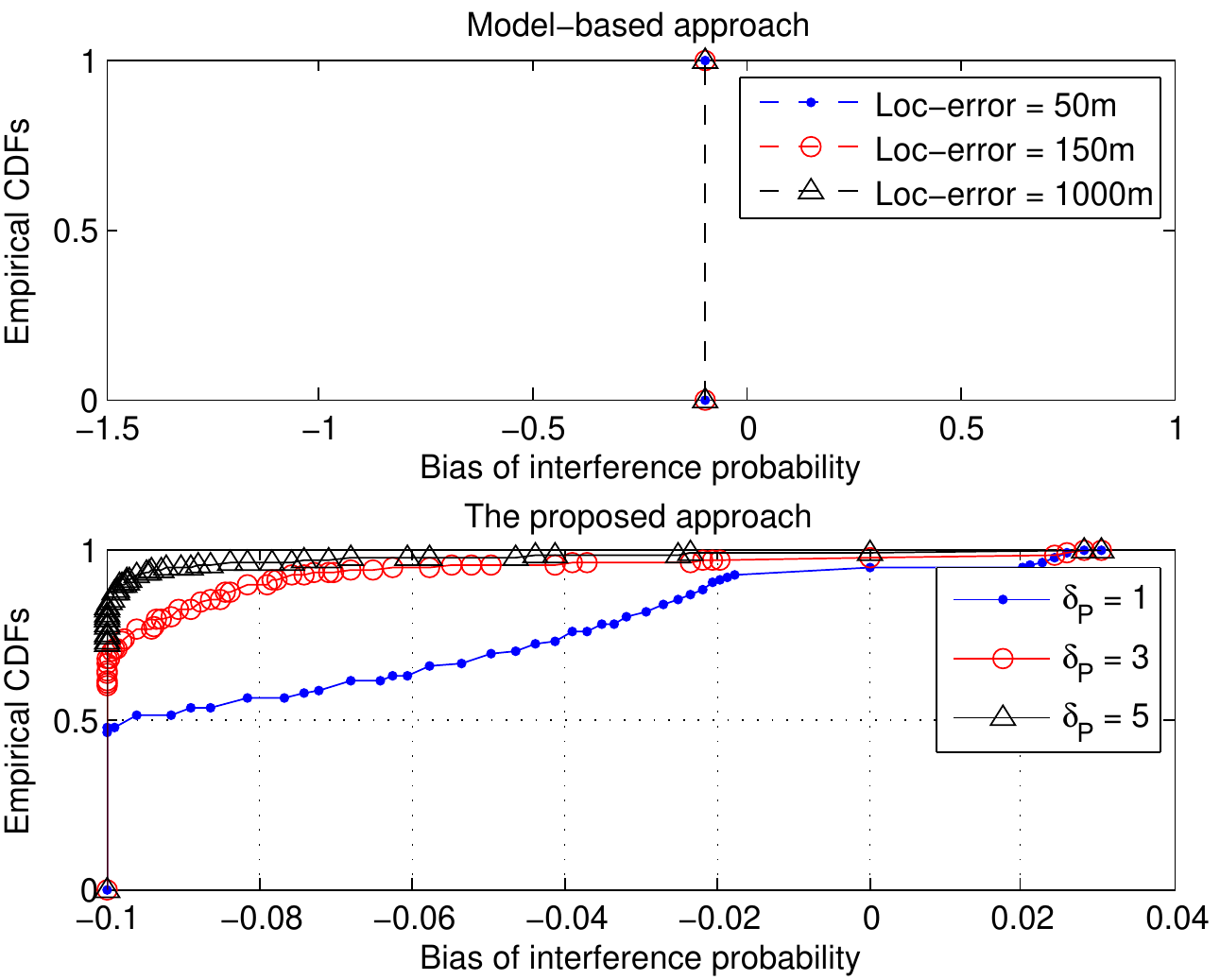}
\label{fig:stacksub:a}}
\caption{Simulation results of \textbf{Scenario II} (Part II).}
\label{fig_CaseII_MIFTP_IP}
\end{figure*}

Fig.~\ref{fig_Sim_Case_II} and Fig.~\ref{fig_CaseII_MIFTP_IP} shows the simulation results under \textbf{Scenario II}. As shown in Fig.~\ref{fig_Sim_Case_II}(a), in this simulation, we consider the case that a cellular BS located at (119.2,0), which is inside the DTV protection region. The shaded disc area in Fig.~\ref{fig_Sim_Case_II}(a) (around the centered cellular BS) represents the cell range of that cellular BS. There is a shadowing zone due to signal attenuation resulted from large obstacles. As an illustrative instance, the signal attenuation from $x=-116.2$ km to $x=-120.2$ km linearly decreases from -20 dB to 0 dB. Comparing with the Scenario I, this scenario is much more complex since the TVWS is located inside the DTV protection region and the coverage boundary is much more irregular. Intuitively, through comparing Fig.~\ref{fig_Sim_Case_II}(d) and Fig.~\ref{fig_Sim_Case_II}(a), it seems the proposed approach performs well.

Moreover, it can be observed in Fig.~\ref{fig_CaseII_MIFTP_IP} that:
\begin{itemize}
  \item In the traditional model-based approach, the biases of IPs for all setups of the localization error are -0.1, which means that all the derived IPs in the model-based approach is zero, considering the interference probability threshold $\nu_{\rm{int}}=0.1$. In other words, all the derived MPEPs are zero and no transmission is allowed in this scenario, which results in a waste of spatial reuse opportunities of TVWS for D2D communications.
  \item Differently, the proposed approach can enable D2D communications in this kind of TVWS while satisfying the interference constraint from the licensed DTV services at more than $90\%$ locations inside the cell for all the three setups of the offset parameter $\delta_P$ (defined in Eq.(\ref{eq:deltaP})). Furthermore, a larger $\delta_P$ corresponds to smaller interference to potential DTV receptions and larger biases of MPEPs. Under this simulation scenario, $\delta_P = 3$ is a better choice to balance the tradeoff between the protection to DTV receptions and the spatial reuse opportunities of D2D communications.
\end{itemize}

To our best knowledge, this is the first try to explore and exploit TVWS inside the DTV protection region. Potential application scenarios include (but not limited to) communications between internet of vehicles in the underground parking, D2D communications in hotspots such as subwaym game stadiums, and airports, etc.

\section{Related Work}
During the past few years, the idea of enabling D2D communications in cellular networks for handling local traffic has gained growing attention. The prior studies in~\cite{D2D_2008,D2D_2009,D2D_Mag,D2D_WMag} have shown that better resource utilization can be achieved by non-orthogonal spectrum sharing between D2D communications and cellular networks. Among many others, the authors of~\cite{D2D_Mutual_Inter2,D2D_Mutual_Inter3,D2D_SLY,D2D_App} have proposed various interference management schemes to coordinate D2D and cellular users for achieving improved spatial reuse of the cellular spectrum. The work of this paper is complementary to prior studies and the goal here is to explore unlicensed TV spectrum for D2D communications with the assistance from cellular networks. The sharing of TV spectrum between D2D and DTV users brings unique challenges, mainly due to the lack of explicit signaling cooperation.

The idea of opportunistic access of TVWS for cellular networks has also received increasing interests. A related work in~\cite{cognitive-radio-enabled} proposed a spectrum sensing-based mechanism to explore TVWS for cellular users, where a simplified circular DTV coverage model was assumed and collaborative sensing was done among neighboring cellular BSs with fixed topology. In contrast, our proposed approach introduces mobile crowd sensing to collect spectrum measurements from massive personal devices and exploits TVWS in a much finer granularity by considering the practical irregular DTV coverage.

Based on the guideline provided by FCC~\cite{FCC-Rule}, several TVWS database systems have been developed by companies~\cite{MicrosoftWhiteSpace,GoogleWhiteSpace,SpectrumBridgeWhiteSpace,Telcordia} and very recently, many of them have entered the phase of public testing~\cite{FCC-TVWS}. Technically, in~\cite{Senseless_2012}, the authors presented a database system, named \emph{SenseLess}, in which unlicensed devices rely on a geolocation database service to determine TVWS availability without spectrum sensing. One common feature of these database systems is that they are \emph{model-based} approaches in essence, since the database service is provided by using a combination of sophisticated signal propagation modeling, fine terrain data, and an up-to-date parameters of the DTV transmitters. Alternatively, this paper presents a \emph{data-driven} approach to build a database by learning the TVWS availability from big spectrum measurement data.

As opposed to propagation model estimates, there are few related work that use actual measurements to build spatial TVWS maps or database. One representative research work has been done by the European project FARAMIR (during 2010-2012)~(see, e.g.,~\cite{Faramir-Deliverables,TVWS_Europe,Measurement-london,REM-MassiveData}), where extensive spectrum measurements have been conducted at several locations in Europe to provide a valuable basis for spectrum use modeling in time, frequency, and space, and to increase the radio environmental and spectral awareness of future wireless systems. In~\cite{Bounding} and~\cite{survey_PathLoss}, a large set of active measurements have been collected to evaluate the accuracy of propagation models in making radio link predictions, where a conclusion has been reached that these models can be used for nationwide coverage planning, but perform poorly at predicting accurate path loss even in relatively simple outdoor environments and more complex models that consider a larger number of variables (e.g., terrain, climactic, soil conductivity, etc) do not necessarily make better predictions. These studies reinforce the motivation of this paper, which extends those studies by developing effective data mining algorithms to build TVWS database from actual measurements.

\section{Conclusions}

In this paper, we proposed a mobile crowd sensing-driven geolocation spectrum database for device-to-device communications (D2D) in TV white space. This paper represents the first try to highlight that both the distance-based path loss and the location-dependent shadowing effect can improve the isolation between the small-scale D2D communications and large-scale DTV services, which increases the flexibility in local spectrum usage and benefits the D2D communications by providing more TVWS opportunities. We formulated the spatial reuse of a TV channel between licensed DTV services and unlicensed D2D communications as an optimization problem. To obtain an effective solution, we presented a systematic approach consisting of mobile crowd sensing for spectrum measurements collection, fast matrix completion algorithm for unknown measurements recovery, nonlinear support vector machine algorithm for irregular coverage boundary, and opportunistic spatial reuse algorithm for deciding the maximum permitted emission power (MPEP) for each D2D link. One interesting but critical research direction is to investigate the coexistence of multiple co-channel D2D links, where mutual interference exists among different D2D links and the total interference reaching a TV receiver is the cumulative sum of received signal power from all D2D links in its vicinity. Another future work is to consider adjacent-channel interference limit required by licensed DTV services, where the correlation in frequency domain and the channel frequency selectivity as studied in~\cite{frequencyselective1,frequencyselective2} should be carefully investigated.

\appendices
\section{Matrix Shrinkage Operator $S_{\nu}(\cdot)$}
Assume $\textrm{M}\in \mathbb{R}^{p \times m}$ and its singular value decomposition is given by $\textrm{M} = \textrm{U}\rm{diag}(\sigma)\textrm{V}^T$, where $\textrm{U}\in \mathbb{R}^{p \times r}, \sigma \in \mathbb{R}^{p}_{+}$, and $\textrm{V}\in \mathbb{R}^{m \times r}$. Given a scalar $\nu>0$, $S_{\nu}(\cdot)$ is defined as
\begin{align}
S_{\nu}(\textrm{M}):={\textrm{U}} \rm{diag}(s_{\nu}(\sigma)) {\textrm{V}}^T
\end{align}
with the vector $s_{\nu}(\sigma)$ defined as
\begin{align}
s_{\nu}(\sigma):=\max\{\sigma-\nu,0\}, \textrm{component~wise}.
\end{align}

Simply speaking, $s_{\nu}(\sigma)$ reduces every nonnegative singular value of $\textrm{M}$ by $\nu$; if any singular value is smaller than $\nu$, it is reduced to zero.

\section*{Acknowledgments}
This work was supported by the National Natural Science Foundation of China under Grant No. 61301160 and No. 61172062.

\ifCLASSOPTIONcaptionsoff
  \newpage
\fi

\begin{IEEEbiography}[{\includegraphics[width=1in,height=1.25in,clip,keepaspectratio]{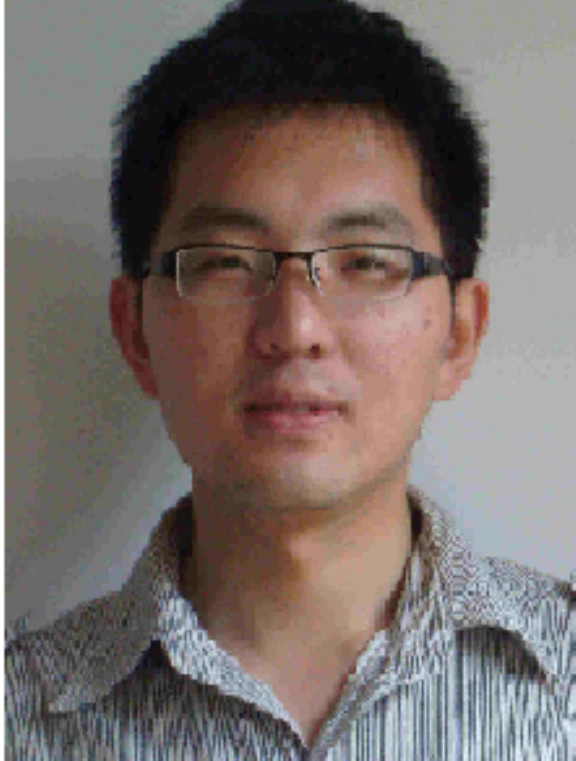}}]{Guoru Ding}
received his B.S. degree in electrical engineering from Xidian University, Xi¡¯an, China, in 2008 and his Ph.D. degree in communications and information systems in College of Communications Engineering, Nanjing, China, in 2014. Since 2014, he has been an assistant professor in College of Communications Engineering, and also a research fellow in National High Frequency Communications Research Center of China. Since April 2015, he has been a Postdoctoral Research Associate at the National Mobile Communications Research Laboratory, Southeast University, Nanjing, China, with Prof. Xiqi Gao. His current research interests include Massive MIMO, Cognitive Radio Networks, Wireless Security, Statistical Learning, and Big Spectrum Data Analytics for 5G Wireless Networks.

He currently serves as an Editor of KSII Transactions on Internet and Information Systems, and an invited reviewer for 10+ Journals such as IEEE Signal Processing Magazine, IEEE Communications Magazine, IEEE Transactions on Signal Processing, IEEE Transactions on Communications, and IEEE Transactions on Wireless Communications, etc. He was a recipient of Best Paper Awards from IEEE VTC2014-Fall and IEEE WCSP 2009. He is a IEEE/ACM/CCF member, the Secretary of IEEE 1900.6 Standard Association Working Group, and also a voting member of IEEE 1900.6.
\end{IEEEbiography}

\begin{IEEEbiography}[{\includegraphics[width=1in,height=1.25in,clip,keepaspectratio]{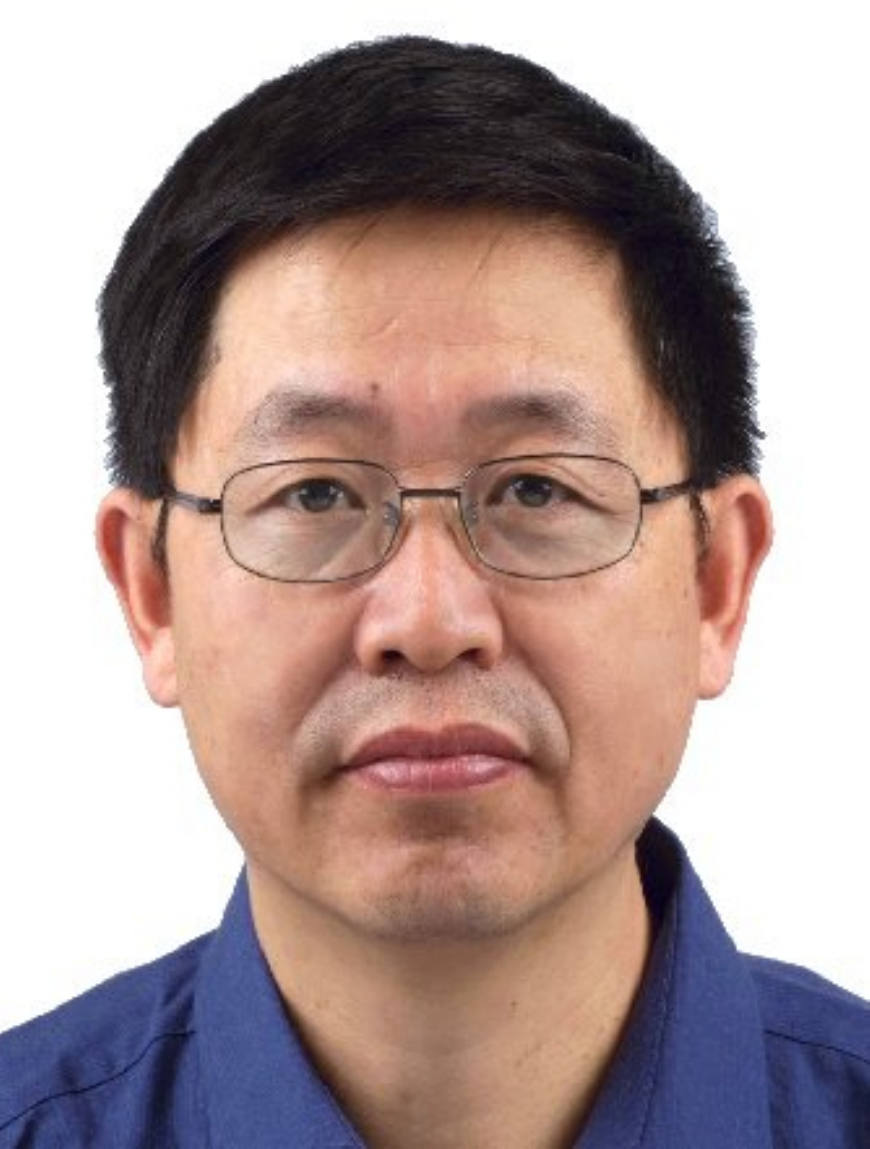}}]{Jinlong Wang}
received his B.S. degree in wireless communications, M.S. degree and Ph.D. degree in communications and electronic systems from Institute of Communications Engineering, Nanjing, China, in 1983, 1986 and 1992, respectively.

He is currently a Chair Professor at PLA University of Science and Technology, Nanjing, China. He was also the co-chair of IEEE Nanjing Section. He has published widely in the areas of signal processing for wireless communications and networking. His current research interests include soft defined radio, cognitive radio, and green wireless communication systems.
\end{IEEEbiography}

\begin{IEEEbiography}[{\includegraphics[width=1in,height=1.25in,clip,keepaspectratio]{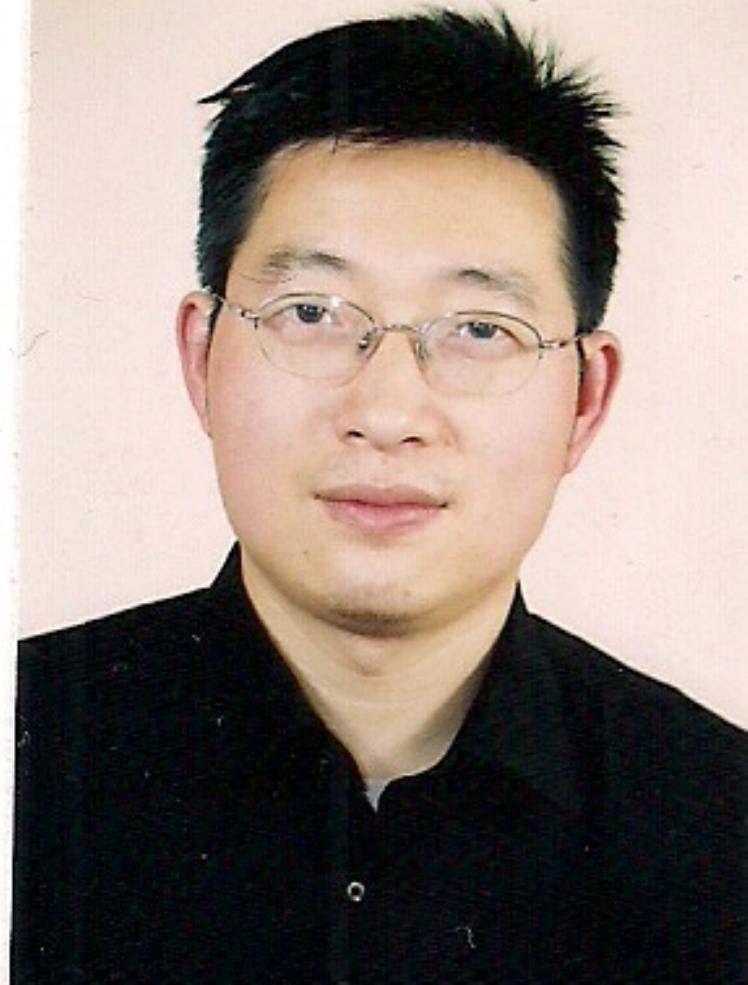}}]{Qihui Wu}
received his B.S. degree in communications engineering, M.S. degree and Ph.D. degree in communications and information systems from Institute of Communications Engineering, Nanjing, China, in 1994, 1997 and 2000, respectively. From 2003 to 2005, he was a Postdoctoral Research Associate at Southeast University, Nanjing, China. From 2005 to 2007, he was an Associate Professor with the Institute of Communications Engineering, PLA University of Science and Technology, Nanjing, China, where he is currently a Full Professor. From March 2011 to September 2011, he was an Advanced Visiting Scholar in Stevens Institute of Technology, Hoboken, USA.

Dr. Wu's current research interests span the areas of wireless communications and statistical signal processing, with emphasis on system design of software defined radio, cognitive radio, and smart radio.
\end{IEEEbiography}

\begin{IEEEbiography}[{\includegraphics[width=1in,height=1.25in,clip,keepaspectratio]{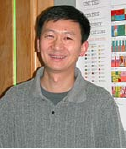}}]{Yu-Dong Yao}
(S'88-M'88-SM'94-F'11) received the B.Eng. and M.Eng. degrees from Nanjing University of Posts and Telecommunications, Nanjing, China, in 1982 and 1985, respectively, and the Ph.D. degree from Southeast University, Nanjing, China, in 1988, all in electrical engineering.
He has been with Stevens Institute of Technology, Hoboken, New Jersey, since 2000 and is currently a professor and department director of electrical and computer engineering. He is also a director of Stevens¡¯ Wireless Information Systems Engineering Laboratory (WISELAB). Previously, from 1989 and 1990, he was at Carleton University, Ottawa, Canada, as a Research Associate working on mobile radio communications. From 1990 to 1994, he was with Spar Aerospace Ltd., Montreal, Canada, where he was involved in research on satellite communications. From 1994 to 2000, he was with Qualcomm Inc., San Diego, CA, where he participated in research and development in wireless code-division multiple-access (CDMA) systems. He holds one Chinese patent and twelve U.S. patents. His research interests include wireless communications and networks, spread spectrum and CDMA, antenna arrays and beamforming, cognitive and software defined radio (CSDR), and digital signal processing for wireless systems.
\end{IEEEbiography}

\begin{IEEEbiography}[{\includegraphics[width=1in,height=1.25in,clip,keepaspectratio]{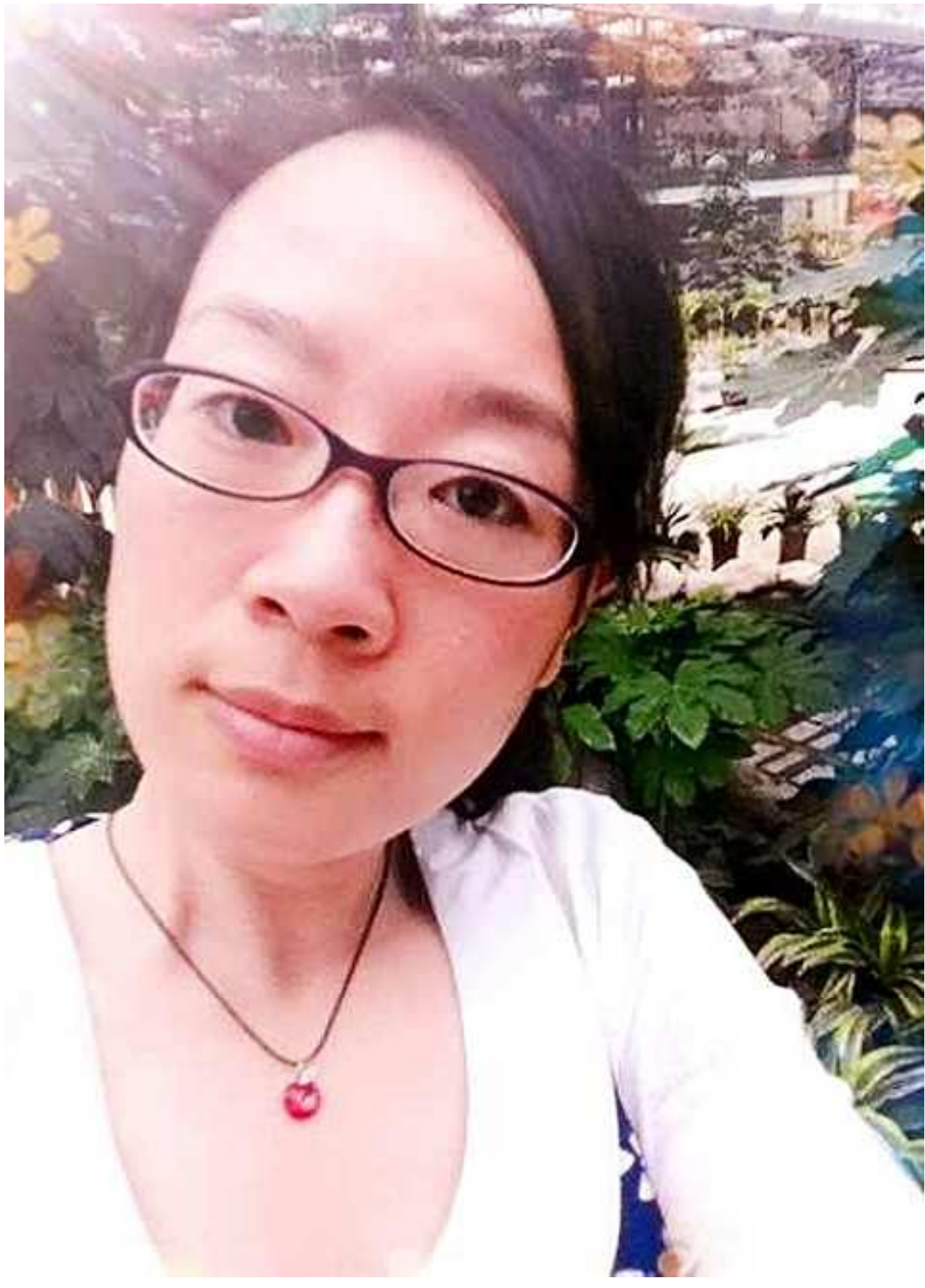}}]{Fei Song}
received her B.S. degree in communications engineering, and her Ph.D. degree in communications and information system from Institute of Communications Engineering, PLA University of Science and Technology, Nanjing, China, in 2002 and 2007, respectively. She is currently an associate professor of PLA University of Science and Technology. Her current research interests are cognitive radio networks, MIMO and statistical signal processing.
\end{IEEEbiography}

\begin{IEEEbiography}[{\includegraphics[width=1in,height=1.25in,clip,keepaspectratio]{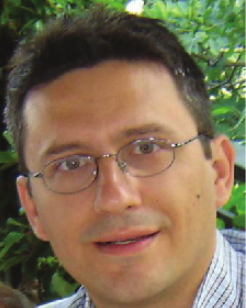}}]{Theodoros A. Tsiftsis}
(S'02-M'04-SM'10) was born in Lamia, Greece, in 1970. He received the B.S. degree in physics from the Aristotle University of Thessaloniki, Thessaloniki, Greece, in 1993; the M.Sc. degree in digital systems engineering from the Heriot-Watt University, Edinburgh, U.K., in 1995; the M.Sc. degree in decision sciences from the Athens University of Economics and Business, Athens, Greece, in 2000; and the Ph.D. degree in electrical engineering from the University of Patras, Patras, Greece, in 2006. He is currently an Assistant Professor with the Department of Electrical Engineering, Technological Educational Institute of Central Greece, Lamia, Greece, and also a collaborative researcher with the Wireless Communications Systems Group (WCSG), Department of Electrical $\&$ Computer Engineering, Aristotle University of Thessaloniki, Thessaloniki, Greece. He is the author of more than 70 technical papers in scientific journals and international conferences. His research interests include the broad areas of cooperative communications, communication theory, wireless communications, and optical wireless communication systems.

Dr. Tsiftsis acts as reviewer for several international journals and he is member of the Editorial Boards of IEEE TRANSACTIONS ON COMMUNICATIONS and IEEE COMMUNICATIONS LETTERS.
\end{IEEEbiography}

\end{document}